\documentclass[12pt]{article} 
\usepackage[utf8]{inputenc} 

\usepackage{graphicx} 
\usepackage{amsmath, amsfonts, amssymb, color, amsxtra}
\usepackage{bm}
\usepackage{enumerate}
\usepackage{natbib}
\usepackage{lineno}
\usepackage[utf8]{inputenc}
\usepackage{verbatim}
\usepackage{appendix}
\usepackage{amsmath}
\usepackage{amsfonts}
\usepackage{amssymb}
\usepackage{amsthm}
\usepackage{graphicx}
\usepackage{natbib}
\usepackage{color}
\usepackage{caption}
\usepackage{subcaption}
\usepackage[affil-it]{authblk}
\usepackage{algpseudocode, algorithm, algorithmicx}

\usepackage[top=1in, bottom=1in, left=0.8in, right=0.8in]{geometry} 
\usepackage{graphicx} 
\usepackage{amsmath, amsfonts, color}
\usepackage{bm}
\usepackage{enumerate}

\usepackage{booktabs} 
\usepackage{array} 
\usepackage{paralist} 
\usepackage{verbatim} 
\usepackage{subfig} 

\usepackage{fancyhdr} 
\usepackage{sectsty}
\allsectionsfont{\sffamily\mdseries\upshape} 

\usepackage[nottoc,notlof,notlot]{tocbibind} 
\usepackage[titles,subfigure]{tocloft} 

\usepackage{setspace}
\newtheorem{theorem}{Theorem}[section]
\newtheorem{lemma}[theorem]{Lemma}

\newtheorem{definition}[theorem]{Corollary}

\newenvironment{remark}[1][Remark]{\begin{trivlist}
\item[\hskip \labelsep {\bfseries #1}]}{\end{trivlist}}

\newcommand{\bx}{{\bm x}}
\newcommand{\bX}{{\bm X}}
\newcommand{\by}{{\bm y}}

\newcommand{\bL}{{\bm L}}

\newcommand{\bz}{{\bm z}}

\newcommand{\bgamma}{{\bm \gamma}}
\newcommand{\bbeta}{{\bm \beta}}

\newcommand{\bSigma}{{\bm \Sigma}}

\newcommand{\bepsilon}{{\bm \epsilon}}

\newcommand{\bzero}{{\bm 0}}

\newcommand{\bI}{{\bm I}}
\newcommand{\bA}{{\bm A}}
\newcommand{\bB}{{\bm B}}

\newcommand{\E}{\mathrm{E}}
\newcommand{\Var}{\mathrm{Var}}
\newcommand{\Cov}{\mathrm{Cov}}
\DeclareMathOperator*{\argmin}{arg\,min}
\providecommand{\keywords}[1]{\textbf{\textit{Keywords---}} #1}

\newcommand\smallO{
  \mathchoice
    {{\scriptstyle\mathcal{O}}}
    {{\scriptstyle\mathcal{O}}}
    {{\scriptscriptstyle\mathcal{O}}}
    {\scalebox{.7}{$\scriptscriptstyle\mathcal{O}$}}
  }

\date{} 

\author{Sen Zhao\thanks{senz@u.washington.edu.}} 
\author{Ali Shojaie\thanks{ashojaie@u.washington.edu.}}
\affil{University of Washington}

\title{A Significance Test for Graph-Constrained Estimation}
\begin{document}

\maketitle
\def\spacingset#1{\renewcommand{\baselinestretch}%
{#1}\normalsize} \spacingset{1}
\abstract{
Graph-constrained estimation methods encourage similarities among neighboring covariates presented as nodes on a graph, which can result in more accurate estimations, especially in high dimensional settings. Variable selection approaches can then be utilized to select a subset of variables that are associated with the response. However, existing procedures do not provide measures of uncertainty of the estimates. Moreover, the vast majority of existing approaches assume that available graphs accurately capture the association among covariates; violating this assumption could severely hurt the reliability of the resulting estimates. In this paper, we present an inference framework, called the Grace test, which simultaneously produces coefficient estimates and corresponding $p$-values while incorporating the external graph information. We show, both theoretically and via numerical studies, that the proposed method asymptotically controls the type-I error rate regardless of the choice of the graph. 
When the underlying graph is informative, the Grace test is asymptotically more powerful than similar tests that ignore external information. We further propose a more general Grace-ridge test that results in a higher power than the Grace test when the choice of the graph is not fully informative. Our numerical studies show that as long as the graph is reasonably informative, the proposed testing methods deliver improved statistical power over existing inference procedures that ignore external information.
}


\spacingset{1.5}

\section{Introduction}\label{sec:itro}
Interactions among genes, proteins and metabolites shed light into underlying biological mechanisms, and clarify their roles in carrying out cellular functions \citep{Zhuetal2007, Michailidis2012}. 
This has motivated the development of many statistical methods to incorporate existing knowledge of biological networks into data analysis \citep[see e.g.][]{Kongetal2006, WeiPan2008, ShojaieMichailidis2009, ShojaieMichailidis2010}. 
 Such methods can lead to identification of novel biological mechanisms associated with the onset and progression of complex diseases \citep[see e.g.][]{Khatrietal2012}.

External network information may be summarized using an undirected weighted graph $G=(V,E,W)$, whose node set $V=\{1,...,p\}$ corresponds to $p$ covariates. 
The edge set $E$ of the graph encodes similarities among covariates, in the sense that two vertices $u,v\in V$ are connected with an edge $e=(u\sim v)\in E$ if covariates $u$ and $v$ are ``similar'' to each other. 
The similarity between neighboring nodes ($u\sim v$) is captured by weights $w(u, v)$. 
Such similarities can for instance correspond to interactions between genes or phylogenetic proximities of species.
%

A popular approach for incorporating network information is to encourage smoothness in coefficient estimates corresponding to neighboring nodes in the network using a \emph{network smoothing penalty} \citep{LiLi2008, Slawskietal2010, Panetal2010, LiLi2010, Huangetal2011, Shenetal2012}. This approach can also be generalized to induce smoothness among similar covariates defined based on a distance matrix or ``kernel'' \citep{Randolphetal2012} which, for instance, capture similarities among microbial communities according to lineages of a phylogenetic tree \citep{Fukuyamaetal2012}. 

The smoothness induced by the network smoothing penalty can result in more accurate parameter estimations, particularly when the sample size $n$ is small compared to the number of covariates $p$. 
Sparsity-inducing penalties, like the $\ell_1$ penalty \citep{LiLi2008,LiLi2010} or the minimum convex penalty (MCP) \citep{Huangetal2011}, can then be used to select a subset of covariates $\bX$ associated with the response $\by$ for improved interpretability and reduced variability. 
It has been shown that, under appropriate assumptions, the combination of network smoothing and sparsity-inducing penalties can consistently select the subset of covariates associated with the response \citep{Huangetal2011}. 
However, such procedures do not account for the uncertainty of the estimator, and in particular, do not provide $p$-values.  

A number of new approaches have recently been proposed for formal hypothesis testing in penalized regression, including resampling and subsampling approaches \citep{MeinshausenBuhlmann2010}, ridge test with deterministic design matrices \citep{Buhlmann2013}, and the low-dimensional projection estimator (LDPE) for $\ell_1$-penalized regression \citep{ZhangZhang2014, vandeGeeretal2014}. 
However, there are currently no inference procedures available for methods that incorporate external information using smoothing penalties. Inference procedures for kernel machine learning methods \citep{Liuetal2007}, on the other hand, test the global association of covariates and are hence not appropriate for testing the association of individual covariates.

Another limitation of existing approaches that incorporate external network information, including those using network smoothing penalties, is their implicit assumption that the network is accurate and informative. However, existing networks may be incomplete or inaccurate \citep{Hartetal2006GB}. As shown in \citet{ShojaieMichailidis2010network}, such inaccuracies can severely impact the performance of network-based methods. Moreover, even if the network is accurate and complete, it is often unclear whether network connectivities correspond to similarities among corresponding coefficients, which is necessary for methods based on network smoothing penalties.  

To address the above shortcomings, we propose a testing framework, the \emph{Grace test}, which incorporates external network information into high dimensional regression and corresponding inferences. The proposed framework builds upon  the graph-constrained estimation (Grace) procedure of \citet{LiLi2008}, \citet{Slawskietal2010} and \citet{LiLi2010}, and utilizes recent theoretical developments for the ridge test by \citet{Buhlmann2013}. 
As part of our theoretical development, we generalize the ridge test with fixed design to the setting with random design matrices $\bX$. 
This generalization was suggested in the discussion of \citet{Buhlmann2013} as a possible extension of the ridge test, and results in improved power compared to the original proposal.

Our theoretical analysis shows that the proposed testing framework controls the type-I error rate, regardless of the informativeness and accuracy of the incorporated network. We also show, both theoretically and using simulation experiments, that if the network is accurate and informative, the Grace test offers improved power over existing approaches that ignore such information. Finally, We propose an extension of the Grace test, called the Grace-ridge or \textit{GraceR} test, for settings where the network may be inaccurate or uninformative. 

The rest of the paper is organized as follows. In Section~\ref{sec:grace}, we introduce the Grace estimation procedure and the Grace test. We also formally define the ``informativeness'' of the network. 
Section~\ref{sec:power} investigates the power of the Grace test, in comparison to its competitors. 
In Section~\ref{sec:gracer}, we propose the Grace-ridge (GraceR) test for robust estimation and inference with potentially uninformative networks. We apply our methods to simulated data in Section~\ref{sec:sim} and to data from The Cancer Genome Atlas (TCGA) in Section~\ref{sec:rd}. We end with a discussion in Section~\ref{sec:disc}. 
Proofs of theoretical results and additional details of simulated and real-data analyses are gathered in Section~\ref{sec:supp}.

Throughout this paper, we use normal lowercase letters to denote scalars, bold lowercase letters to denote vectors and bold uppercase letters to denote matrices. We denote columns of an $n \times p$ matrix $\bX$ by $\bx_j, j = 1,..., p$ and its rows by $\bx^i, i = 1,..., n$. For any two symmetric matrices $\bA$ and $\bB$, we denote $\bA\preceq\bB$ if $\bB-\bA$ is positive semi-definite, or $\lambda_0(\bB-\bA)\geq0$, where $\lambda_0$ denotes the smallest eigenvalue of a symmetric matrix. For an index set $J$, we denote by $\bA_{(J, J)}$ the $|J|\times|J|$ sub-matrix corresponding to the rows and columns indexed by $J$. Finally, for a $p$-vector $\bbeta$, we let $\|\bbeta\|_k\triangleq(\sum_{i = 1}^p|\beta_i|^k)^{1/k}$ for $k \in\mathbb{Z}^+$ and $\|\bbeta\|_\infty\triangleq\max_i \beta_i$. 

\section{The Grace Estimation Procedure and the Grace Test}
\label{sec:grace}

\subsection{The Grace Estimation Procedure}
\label{sec:graceest}

Let $\bL$ be the matrix encoding the external information in an undirected weighted graph $G=(V,E,W)$. 
In general, $\bL$ can be any positive semi-definite matrix, or kernel, capturing the ``similarity'' between covariates. In this paper, however, we focus on the case where $\bL$ is the graph Laplacian matrix, 
\[
 \bL_{(u, v)}\triangleq \left\{
     \begin{array}{ll}
         d_u & \text{if } u = v \\
       -w(u, v) & \text{if } u \text{ and } v \text{ are connected}\\
       0 &  \text{otherwise}
     \end{array}
   \right.,
\]
with $d_u = \sum_{v\sim u} w(u, v)$ denoting the degree of node $u$. 
We also assume that weights $w(u, v)$ are nonnegative. However, the definition of Laplacian and the analysis in this paper can be generalized to also accommodate negative weights \citep{chung1997spectral}. 

Let $\bX=(\bx_1, ...,\bx_p)\in\mathbb{R}^{n\times p}$ be the $n \times p$ design matrix and $\by\in\mathbb{R}^n$ be the response vector in the linear model 
\begin{equation}\label{model}
\by = \bX\bbeta^\ast + \bepsilon, \hspace{0.5cm} 
\bepsilon \sim N_n(\bzero, \sigma_\bepsilon^2\bI_n), \hspace{0.5cm} 
\bx^i \sim^{iid} N_p(\bzero,\bSigma) \text{ for } i = 1,..., n.
\end{equation}
Multivariate normality of covariates is commonly assumed in analysis of biological networks, particularly, when estimating interactions among genes or proteins using Gaussian graphical models \citep[see e.g.][]{delaFuenteetal2004}. Interestingly, the underlying assumption of network smoothing penalties -- that connected covariates after scaling have similar associations with the response -- is also related to the assumption of multivariate normality \citep{ShojaieMichailidis2010}.
Without loss of generality, we assume $\by$ is centered and columns of $\bX$ are centered and scaled, i.e. 
$\sum_{i=1}^ny_i = 0$ and $\sum_{i=1}^n X_{(i, j)} = 0$, $\bx_j^\top\bx_j=n$ for $j=1,...,p$.
We denote the scaled Gram matrix by $\hat\bSigma\triangleq\bX^\top\bX/n$.

For a non-negative tuning parameter $h$, Grace solves the following optimization problem:
\begin{align}
\label{eq:grace}
\hat\bbeta(h) = \argmin_\bbeta \left\{\big\|\by-\bX\bbeta\big\|_2^2 + h\bbeta^\top\bL\bbeta\right\} = \big(n\hat\bSigma+h\bL\big)^{-1}\bX^\top\by.
\end{align}
When $\bL$ is the Laplacian matrix, $\bbeta^\top\bL\bbeta = \sum_{u\sim v} (\bbeta_u - \bbeta_v)^2 w(u,v)$ \citep{Huangetal2011}. Hence, the Grace penalty $\bbeta^\top\bL\bbeta$ encourages smoothness in coefficients of connected covariates, according to weights of edges. Henceforth, we call $\bL$ the penalty weight matrix. 

For any tuning parameter $h>0$, Equation (\ref{eq:grace}) will have a unique solution if $(n\hat\bSigma + h\bL)$ is invertible.
However, if $p > n$ and $rank(\bL) < p$ this condition may not hold. 
With a Gaussian design $\bx^i \sim^{iid} N_p(\bzero,\bSigma)$, it follows from \citet{Bai1999} that if $\liminf_{n\to\infty}\lambda_0(\bSigma)>0$, and if there exists a sequence of index sets $C_n\subset\{1,...,p\}$, $\lim_{n\to\infty}|C_n|/n<1$, such that $\liminf_{n\to\infty}\lambda_0(\bL_{(V\backslash C_n, V\backslash C_n)})>0$, then $(n\hat\bSigma+h\bL)$ is almost surely invertible. In this section we hence  assume that $(n\hat\bSigma+ h\bL)$ is invertible. This condition is relaxed in Section~\ref{sec:gracer}, when we propose the more general Grace-ridge (GraceR) test.


As mentioned in the Introduction, several methods have been proposed to select the subset of relevant covariates for Grace. For example, \citet{LiLi2008,LiLi2010} added an $\ell_1$ penalty to the Grace objective function,
\begin{align}
\hat{\bbeta}_{\ell_1}(h, h_1) = \argmin_\bbeta \left\{\big\|\by-\bX\bbeta\big\|_2^2  + h\bbeta^\top\bL\bbeta + h_1\big\|\bbeta\big\|_1\right\}.
\end{align}
\citet{Huangetal2011} instead added the MCP and proposed the sparse Laplacian shrinkage (SLS) estimator. 
While these methods perform automatic variable selection, they do not provide measures of uncertainty, i.e. confidence intervals or $p$-values. In this paper, we instead propose an inference procedure that  provides $p$-values for estimated coefficients from Equation~\eqref{eq:grace}. The resulting $p$-values can then be used to assess the significance of individual covariates, and select a subset of relevant variables.

\subsection{The Grace Test}
\label{sec:gracetest}

Before introducing the Grace test, we present a lemma that characterizes the bias of the Grace estimation procedure.
\begin{lemma}
\label{lem:bias}
For any $h > 0$, assume $(n\hat\bSigma+h\bL)$ is invertible. Then, given 
$\bX$, $\hat{\bbeta}(h)$ as formulated in \eqref{eq:grace} is an unbiased estimator of $\bbeta^\ast$ if and only if $\bL\bbeta^\ast = \bzero$. Moreover,
\begin{align}\label{eq:gracebias}
\big\|\mathbf{Bias}(\hat{\bbeta}(h)|\bX)\big\|_2&\leq \frac{h\|\bL\bbeta^\ast\|_2}{\lambda_{0}(n\hat\bSigma+h\bL)}. 
\end{align}
\end{lemma}
Because the bias of the Grace estimator depends directly on the magnitude of $\bL\bbeta^\ast$, we consider $\bL$ to be informative if $\bL\bbeta^\ast$ is small. According to Lemma~\ref{lem:bias}, the Grace estimator will be unbiased only if $\bbeta^\ast$ lies in the space spanned by the eigenvectors of $\bL$ with 0 eigenvalues. In reality, however, this condition cannot be checked from data. Thus, to control the type-I error rate, we must adjust for this potential estimation bias. 

Our testing procedure is motivated by the ridge test proposed in \citet{Buhlmann2013}, which we briefly discuss next. 
First, note that ridge is also a biased estimator of $\bbeta^\ast$, and its \emph{estimation bias} is negligible only if the ridge tuning parameter is close to zero. 
In addition to the estimation bias, \citet{Buhlmann2013} also accounted for the \emph{projection bias} of ridge regression for a \textit{fixed} design matrix $\bX$.  This is because for fixed design matrices with $p>n$, $\bbeta^\ast$ is not uniquely identifiable, as there are infinitely many $\bbeta$'s such that $\E(\by)=\bX\bbeta$. Using ridge regression, $\bbeta^\ast$ is only estimable if it lies in the row space of $\bX$, $\mathcal{R}(\bX)$, which is a proper subspace of $\mathbb{R}^p$ when $p>n$. 
If $\bbeta^\ast$ does not lie in this subspace, the ridge estimated regression coefficient is indeed the projection of $\bbeta^\ast$ onto $\mathcal{R}(\bX)$, which is not identical to $\bbeta^\ast$. This gives rise to the projection bias. 

To account for these two types of biases, \citet{Buhlmann2013} proposed to shrink the ridge estimation bias to zero by shrinking the ridge tuning parameter to zero, while controlling the projection bias using a stochastic bias bound derived from a lasso initial estimator. A side effect of shrinking the ridge tuning parameter to zero is that the variance of covariates with high multi-collinearity could become large; this would hurt the statistical power of the ridge test. In addition, the stochastic bound for the projection bias is rather loose. This double-correction of bias further compromises the power of the ridge test.

In this paper, we develop a test for random design matrices, which was suggested in the discussion of \citet{Buhlmann2013} as a potential extension. With random design matrices, we do not incur any projection bias. This is because the regression coefficients in this case are uniquely identifiable as $\bSigma^{-1}\Cov(\bX,\by)$ under the joint distribution of $(\bX, \by)$. Here, $\bSigma$ denotes the population covariance matrix of covariates and $\Cov(\bX,\by)$ is the population covariance between the covariates and the response; see \citet{ShaoDeng2012} for a more elaborate discussion of identifiability for fixed and random design matrices. 

To control the type-I error rate of the Grace test, we adjust for the potential estimation bias using a stochastic bound derived from an initial estimator. By adjusting for the estimation bias using a stochastic upper bound, the Grace tuning parameter needs not be very small. Thus, the variances of Grace estimates are less likely to be unreasonably large; this results in improved power for the Grace test. 
Power properties of the Grace test are more formally investigated in Section~\ref{sec:power}. Next, we formally introduce our testing procedure.

Consider the null hypothesis $H_0: \beta_j^\ast =0$ for some $j\in\{1,...,p\}$. Let $\tilde\bbeta$ be an initial estimator with asymptotic $\ell_1$ estimation accuracy, i.e. $\| \tilde\bbeta - \bbeta^\ast \|_1 = \smallO_p(1)$. 
The Grace test statistic is defined as 
\begin{align}
\hat\bz^G = \hat\bbeta(h) + h(n\hat\bSigma+h\bL)^{-1}\bL\tilde\bbeta,
\end{align}
where $\hat\bbeta(h)$ is the Grace estimator from \eqref{eq:grace} with tuning parameter $h$. Plugging in \eqref{eq:grace} and adding and subtracting $h(n\hat\bSigma+h\bL)^{-1}\bL\tilde\bbeta$, we can write
\begin{align}
\label{eq:keyrel}
\hat z_j^G = \beta_j^\ast + Z_j^G +\gamma_j^G, \qquad j = 1,...,p, 
\end{align}
where 
\begin{align*}
Z_j^G|\bX&\sim N\left( 0, n\sigma_\bepsilon^2 \left[ (n\hat\bSigma+h\bL)^{-1} \hat\bSigma(n\hat\bSigma+h\bL)^{-1} \right]_{(j,j)} \right), \\
\bgamma^G &\triangleq h(n\hat\bSigma+h\bL)^{-1}\bL(\tilde\bbeta-\bbeta^\ast).
\end{align*}
Next, we derive an asymptotic stochastic bound for $\gamma_j^G$ such that under the null hypothesis
\begin{equation}\label{eq:keycon}
|\gamma_j^{G}|\precsim^{asy.}\Gamma_j^G 
\hspace{.1cm} \text{ or equivalently, } \hspace{.1cm}
\lim_{n\to\infty}Pr\left(|\gamma_j^{G}|\leq\Gamma_j^G\right)= 1. 
\end{equation}
Then, under the null hypothesis, $|\hat z^{G}_j|\precsim^{asy.}|Z_j^G|+\Gamma_j^G$, which allows us to asymptotically control the type-I error rate. 

To complete our testing framework, we use the fact under suitable conditions and with proper tuning parameter $h_{Lasso}$, described in Theorem~\ref{thm:stbound}, the $\ell_1$ estimation error of the lasso, 
\begin{align}
\label{eq:lasso}
\tilde\bbeta(h_{Lasso}) = \argmin_\bbeta \left\{\frac{1}{n}\big\|\by-\bX\bbeta\big\|_2^2 + h_{Lasso}\big\|\bbeta\big\|_1\right\},
\end{align}
is asymptotically controlled \citep{BuhlmannvandeGeer2011}. We thus use the lasso estimator as the initial estimator for the Grace test, i.e. $\tilde\bbeta \triangleq \tilde\bbeta(h_{Lasso})$.
Theorem~\ref{thm:stbound} then constructs a $\Gamma_j^G$ that satisfies Condition~\eqref{eq:keycon}. 
First, we present required conditions. 
\begin{itemize}
	\item {\bf{A0}}: $(n\hat\bSigma+h\bL)$ is invertible.
	\item {\bf{A1}}: $\by=\bX\bbeta^\ast+\bepsilon$ where $\bx^i \sim^{iid} N_p(\bzero,\bSigma) \text{ for } i = 1,..., n$ and $\bepsilon\sim N_n(\bzero, \sigma_\bepsilon\bI)$.
	\item {\bf{A2}}: Let $S_0 \triangleq \{j: \beta^\ast_j\neq 0\}$ be the active set of $\bbeta^\ast$ with cardinality $s_0\triangleq|S_0|$. We have $s_0 = \smallO\left( \big[n/\log{p}\big]^\xi \right)$ for some $0<\xi<1/2$.
	\item {\bf{A3}}: The $\bSigma$-compatibility condition \citep{BuhlmannvandeGeer2011} in Definition~\ref{def:compatibility} is met for the set $S_0$ with compatibility constant $\liminf_{n\to\infty}\phi_{\bSigma,n}^2= d>0$, where $d$ is a constant.
	\item {\bf{A4}}: $h$ and $\bL$ are such that 
	\[
	\left[(n\hat\bSigma+h\bL)^{-1}h\bL\right]_{(j,j)} =\mathcal{O}_p\left(\left[\frac{n}{\log{p}}\right]^{\frac{1}{2}-\xi}\right).
	\]
\end{itemize}
\begin{definition}[$\bSigma$-Compatibility Condition]
\label{def:compatibility} 
For an index set $S\subset\{1,...,p\}$ with cardinality $s$, define $\bbeta^S$ and $\bbeta^{S^c}$ such that $\beta_j^S\triangleq\beta_j 1_{\{j\in S\}}$, $\beta_j^{S^c}\triangleq\beta_j 1_{\{j\notin S\}}$. 
We say that the $\bSigma$-compatibility condition is met for the set $S$ with compatibility constant $\phi_\bSigma>0$ if for all $\bbeta\in\mathbb{R}^p$ living in the cone $\|\bbeta^{S^c}\|_1\leq 3\|\bbeta^{S}\|_1$, we have
\begin{align}
\big\|\bbeta^S\big\|_1^2\leq\bbeta^\top\bSigma\bbeta\frac{s}{\phi_\bSigma^2}.
\end{align}
\end{definition}
As discussed in Section~\ref{sec:graceest}, {\bf{A0}} is required for uniqueness of the Grace estimator, and is justified by the Gaussian deign. 
{\bf{A2}} is a standard assumption, and requires the number of relevant covariates to not grow too fast, so that the signal is not substantially diluted among those relevant covariates. Note that with $p=\mathcal{O}\left(\exp(n^\nu)\right)$ for some $\nu<1$,  $s_0$ can grow to infinity as $n\to\infty$. The $\bSigma$-compatibility condition in {\bf{A3}} is closely related to the restricted eigenvalue assumption introduced in \citet{Bickeletal2009}. 
Assumption {\textbf{A4}} is made for improved control of type-I error, and can be relaxed at a cost of potential loss of power with finite samples; see Remark~\ref{rem:A4}.  
On the other hand, given $\bX$ and $\bL$, when $h/n\to\infty$, the eigenvectors and eigenvalues of $(n/h)\hat\bSigma+\bL$ converge to the eigenvectors and eigenvalues of $\bL$. This indicates that $(n\hat\bSigma+h\bL)^{-1}h\bL$ converges to a diagonal matrix with diagonal entries equal to 0 or 1, and {\bf{A4}} is satisfied. 
\begin{theorem}\label{thm:stbound}
Suppose Assumptions {\bf{A0}} -- {\bf{A4}} are satisfied, and let $\tilde\bbeta\triangleq\tilde\bbeta(h_{Lasso})$ with the tuning parameter $h_{Lasso}\asymp\sqrt{\log{p}/n}$. 
Let
\begin{align}
\label{eq:Gamma}
\Gamma_j^G\triangleq h\left\|\big[(n\hat\bSigma+h\bL)^{-1}\bL\big]_{(j,-j)}\right\|_\infty\left(\frac{\log{p}}{n}\right)^{\frac{1}{2}-\xi},
\end{align}
where $\left\|\big[(n\hat\bSigma+h\bL)^{-1}\bL\big]_{(j,-j)}\right\|_\infty \triangleq \max_{i:i\neq j} \big|(n\hat\bSigma+h\bL)^{-1}\bL\big|_{(j,i)}$ is the maximum in absolute value of entries in row $j$ without the diagonal entry.
Then $\Gamma_j^G$ satisfies condition~\eqref{eq:keycon}.

Under the null hypothesis $H_0: \beta_j = 0$, for any $\alpha>0$ we have
\begin{align}
\label{eq:alpha}
\limsup_{n\to\infty} Pr\left(\big|\hat z_j^G\big|>\alpha\right)\leq\limsup_{n\to\infty}\Pr\left(\big|Z_j^G\big|+\Gamma_j^G>\alpha\right).
\end{align}
\end{theorem}
\begin{remark}\label{rem:A4}
If we instead consider 
\[
\Gamma_j^G=h\left\|\big[(n\hat\bSigma+h\bL)^{-1}\bL\big]_{(j,.)}\right\|_\infty\left(\frac{\log{p}}{n}\right)^{\frac{1}{2}-\xi},
\]
we can relax Assumption {\textbf{A4}} and still control the asymptotic type-I error rate. Theorem~\ref{thm:stbound} can then be similarly proved without {\textbf{A4}}. 
However, as $h/n\to\infty$, $(n\hat\bSigma+h\bL)^{-1}h\bL$ converges to a diagonal matrix, in which case $\left\|\big[(n\hat\bSigma+h\bL)^{-1}h\bL\big]_{(j,.)}\right\|_\infty\gg\left\|\big[(n\hat\bSigma+h\bL)^{-1}h\bL\big]_{(j,-j)}\right\|_\infty$. This looser stochastic bound may result in lower power in finite samples. 
\end{remark}

Theorem~\ref{thm:stbound} shows that regardless of the choice of $\bL$, the type-I error rate of the Grace test is asymptotically controlled.  The stochastic bound $\Gamma_j^G$ relies on the unknown sparsity parameter $\xi$.  Following \citet{Buhlmann2013} we suggest a small value of $\xi$, and use $\xi = 0.05$ in the simulation experiments in Section~\ref{sec:sim} and real data example in Section~\ref{sec:rd}. 

Using \eqref{eq:alpha}, we can test $H_0$ using the asymptotically valid two-sided $p$-value
\begin{equation}\label{eq:p}
P_j^G = 2\left(1-\Phi\left[\frac{(|\hat\bz_j^G|-\Gamma_j^G)_+}{\sqrt{\Var(Z_j^G|\bX)}}\right]\right),
\end{equation}
where $\Phi$ is the standard normal c.d.f., and $a_+ = \max(a, 0)$. 
Calculating $p$-values requires estimating $\sigma_\bepsilon^2$ and choosing a suitable tuning parameter $h$. We can estimate $\sigma_\bepsilon^2$ using any consistent estimator, such as the scaled lasso \citep{SunZhang2012}. In the simulation experiments and real data example, we choose $h$ using 10-fold cross-validation (CV).


%

Note that, when simultaneously testing multiple hypotheses: $H_0: \beta_j^\ast = 0$ for any $j\in J\subseteq\{1,...,p\}$ versus $H_a: \beta_j^\ast\neq 0 \text{ for some } j\in J$, we may wish to control the false discovery rate (FDR). Because covariates in the data could be correlated, test statistics on multiple covariates may show arbitrary dependency structure. We thus suggest controlling the FDR using the procedure of \citet{BenjaminiYekutieli2001}. 
Alternatively, we can control the family-wise error rate (FWER) using, e.g. the method of \citet{Holm1979}. 


\section{Power of the Grace Test}
\label{sec:power}

In this section, we investigate power properties of the Grace test. 
Our first result describes sufficient conditions for detection of nonzero coefficients. 
\begin{theorem}\label{thm:suffdet}
Assume Assumptions {\bf{A0}} -- {\bf{A4}} are met. If for some $h$, some $0<\alpha<1$, $0<\psi<1$, conditional on $\bX$, we have
\begin{align}
\label{eq:tuning}
\big|\beta_j^\ast\big| >2\Gamma_j^G + q_{(1-\alpha/2)}\sqrt{\Var(Z_j^G|\bX)} + q_{(1-\psi/2)}, 
\end{align}
where $\Phi\left(q_{(1-\alpha/2)}\right)=1-\alpha/2$. Then using the same tuning parameter $h$ in the Grace test, we get
$
\lim_{n\to\infty} Pr\left(P_j^{G}\leq\alpha\big|\bX\right) \geq \psi.
$
\end{theorem}

Having established the sufficient conditions for detection of non-null hypotheses in Theorem~\ref{thm:suffdet}, we next turn to comparing the power of the Grace test with its competitors: the Grace test, the ridge test with small tuning parameters $h_2=\mathcal{O}(1)$ and no bias correction, and the GraceI test, which is the Grace test with identity penalty weight matrix $\bI$. The ridge test may be considered as a variant of the test proposed in \citet{Buhlmann2013} without the adjustment of the projection bias -- because we assume the design matrix is random, we incur no projection bias in the estimation procedure. 

As indicated in Lemma~\ref{lem:bias}, the estimation bias of the Grace procedure depends on the informativeness of the penalty weight matrix $\bL$. When $\bL$ is informative, we are able to increase the size of the tuning parameter, which shrinks the estimation variance without inducing a large estimation bias. Thus, with an informative $\bL$, we are able to obtain a better prediction performance, as shown empirically in \citet{LiLi2008, Slawskietal2010, LiLi2010}. In such setting, the larger value of the tuning parameter, e.g. as chosen by CV, also results in improved testing power, as discussed next.

Theorem~\ref{thm:power1} compares the power of the Grace test to its competitors in a simple setting of $p=2$ predictors, $\bx_1$ and $\bx_2$. 
In particular, this result identifies sufficient conditions under which the Grace test has asymptotically superior power. It also gives conditions for the GraceI test to have higher power than the ridge test. 
The setting of $p=2$ predictors is considered mainly for ease of calculations, as in this case, we can directly derive closed form expressions of the corresponding test statistics. Similar results are expected to hold for $p>2$ predictors, but require additional derivations and notations.

Assume $\by=\bx_1\beta_1^\ast+ \bx_2\beta_2^\ast+\bepsilon$, where $\bepsilon\sim N_2(\bzero, \sigma^2_\bepsilon\bI)$, and $\bx_1$, $\bx_2$ are scaled. Denote 
\[
\bL \triangleq \left( \begin{array}{ccc}
1 & l \\
l & 1  \end{array} \right), \hspace{1cm}
\hat\bSigma\triangleq\frac{1}{n}\bX^\top\bX= \left( \begin{array}{ccc}
1 & \rho \\
\rho & 1  \end{array} \right).
\]
Theorem~\ref{thm:power1} considers the power for testing the null hypothesis $H_0:\beta_1^\ast=0$, in settings where $\beta_1^\ast\neq0$, without any constraints on $\beta_2^\ast$. 

\begin{theorem}
\label{thm:power1}
Suppose Assumptions {\bf{A0}} -- {\bf{A4}} are met. Let $P_j^{G}(h_n^G)$, $P_j^{GI}(h_n^{GI})$ and $P_j^{R}$ be the Grace, GraceI and ridge $p$-values, respectively, with tuning parameters $h_n^G$ for Grace and $h_n^{GI}$ for GraceI. Define
\begin{align}
\Upsilon_{p,n}(h, l, \rho, |\beta_1|) \triangleq \frac{\left[(h/n+1)^2-(\rho+lh/n)^2\right]\cdot|\beta_1| - [\log{p}/n]^{1/2-\xi}\cdot|(l-\rho)h/n|}{\sqrt{(1+2h/n)(1-\rho^2) + (h/n)^2(1+l^2-2l\rho)}}.
\end{align}
Then, conditional on the design matrix $\bX$, under the alternative hypothesis $\beta_1^\ast =b\neq 0$, the following statements hold with probability tending to 1, as $n\to\infty$.
\begin{itemize}
\item[a)] If $\, \lim\limits_{n\to \infty}\Upsilon_{p,n}(h_n^G, l,\rho, |b|)\geq\lim\limits_{n\to \infty}\Upsilon_{p,n}(h_n^{GI}, 0,\rho, |b|) \,$, then 
	$\,\lim\limits_{n\to\infty} [P_1^{G}(h_n^G)/P_1^{GI}(h_n^{GI})]\leq 1$.
\item[b)] If $\,\lim\limits_{n\to \infty}\Upsilon_{p,n}(h_n^G, l,\rho, |b|)\geq \sqrt{1-\rho^2}\,|b| \,$, then $\,\lim\limits_{n\to\infty} [P_1^{G}(h_n^G)/P_1^{R}] \leq 1$.
\item[c)] If $\,\lim\limits_{n\to \infty}\Upsilon_{p,n}(h_n^{GI}, 0,\rho, |b|)\geq \sqrt{1-\rho^2}\,|b|$\,, then $\,\lim\limits_{n\to\infty} [P_1^{GI}(h_n^{GI})/P_1^{R}] \leq 1$.
\end{itemize}
\end{theorem}

\begin{figure}[h]
\caption{(a) The ratio of $\Upsilon_{p,n}(h_n^G, l,\rho, |\beta_1^\ast|)$ over $\Upsilon_{p,n}(h_n^{GI}, 0,\rho, |\beta_1^\ast|)$ for different $l$ and $\rho$ with $h_n^G / n = h_n^{GI} / n = 10$, $[\log{p}/n]^{1/2-\xi}=0.25$ and $\beta^\ast_1=1$. A plus sign indicates the ratio is greater than 1.02, whereas a minus sign indicates the ratio is smaller than 0.98; filled circles indicate an intermediate value. (b) The log-ratio of $\Upsilon_{p,n}(h_n^G, l,\rho, |\beta_1|)$ over $\sqrt{1-\rho^2}$ for different $l$ and $\rho$ with $h_n^G / n= 10$, $[\log{p}/n]^{1/2-\xi}=0.25$ and $\beta_1^\ast=1$. A plus sign indicates the log-ratio is greater than 0.5 (ratio $>1.65$), whereas a minus sign indicates the log-ratio is smaller than \textsc{-}0.5 (ratio $< 0.61$); filled circles indicate an intermediate value} 
\label{fig:GvO}
\centering
\includegraphics[height = 20cm]{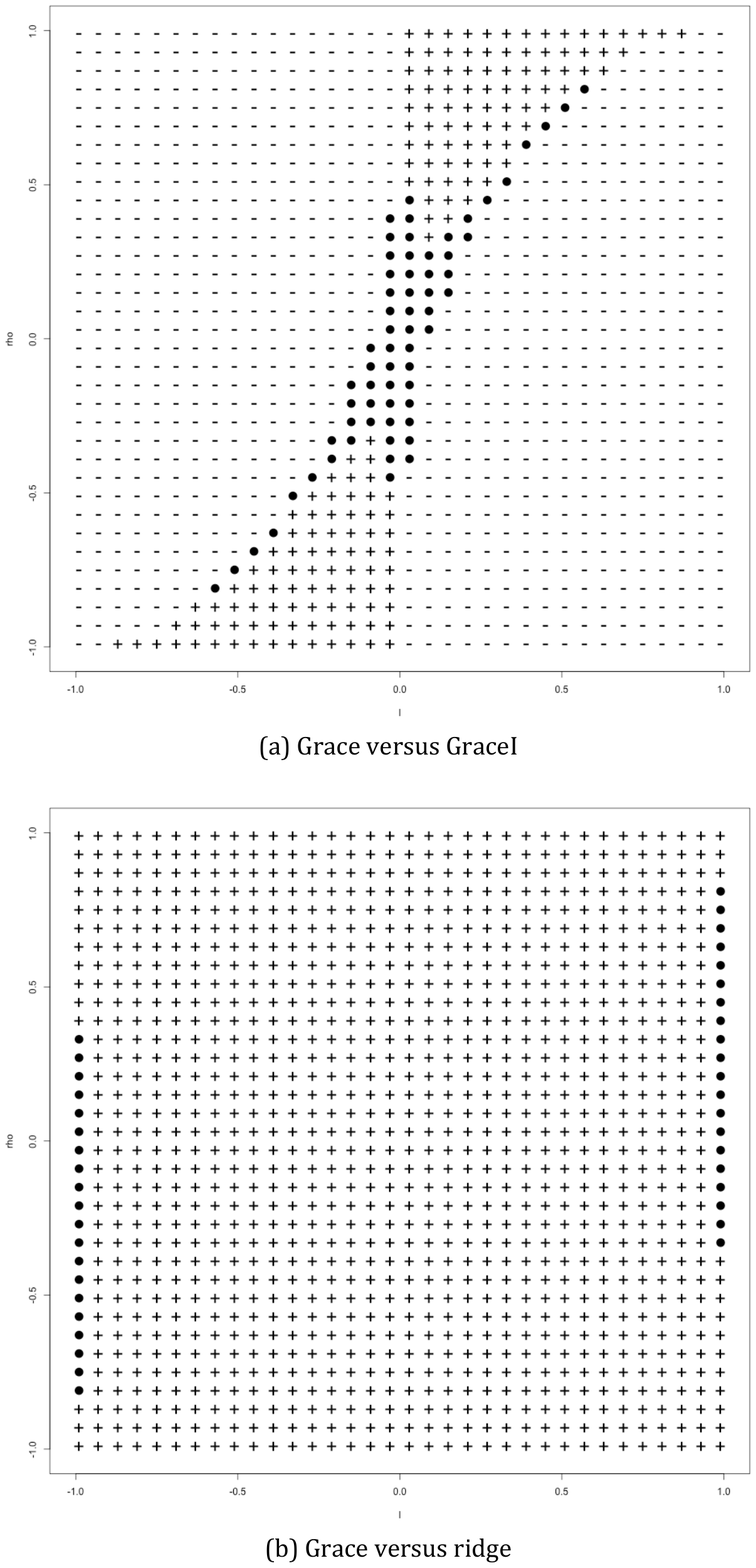}
\end{figure}

Theorem~\ref{thm:power1} indicates that, as $h_n^G / n$ and $h_n^{GI}/n$ diverge to infinity, both $\Upsilon_{p,n}(h_n^G, l,\rho, |\beta_1^\ast|)$ and $\Upsilon_{p,n}(h_n^{GI}, 0,\rho, |\beta_1^\ast|)$  approach infinity. This implies, on one hand, that for $h_n^G$ and $h_n^{GI}$ sufficiently large, both the Grace and GraceI tests are asymptotically more powerful than the ridge test. On the other hand, we can only compare the powers of the Grace and GraceI tests under some constraints on their tuning parameters. 
With equal tuning parameters for Grace and GraceI, $h_n^G=h_n^{GI}$, we can show, after some algebra, that as $h_n^G/n=h_n^{GI}/n\to\infty$,  we have $\lim_{n\to \infty}\Upsilon_{p,n}(h_n^G, l,\rho, |\beta_1^\ast|)\geq\lim_{n\to \infty}\Upsilon_{p,n}(h_n^{GI}, 0,\rho, |\beta_1^\ast|)$ if $(1-l^2) \ge \sqrt{(1+l^2-2l\rho)}$.
In this case, the Grace test is more powerful than the GraceI test if $l$ is between 0 and $l^\ast$, where $l^\ast$ is the unique root  in $[-1,1]$ of the cubic equation $l^3-3l+2\rho=0$. Figure~\ref{fig:GvO}(a) compares the powers of the Grace and GraceI tests with equal tuning parameters $h_n^G / n=h_n^{GI} / n=10$ and $\beta_1^\ast=1$. 
It can be seen that, the Grace test asymptotically outperforms the GraceI test when $l$ is close to $\rho$ with equally large tuning parameters. However, when $\l$ is far from $\rho$, the GraceI test could be more powerful. This observation, and the empirical results in Section~\ref{sec:sim} motivate the development of the GraceR test, introduced in Section~\ref{sec:gracer}. 

A similar comparison for powers of the Grace and the ridge test, with $h_n^G / n=10$ and $\beta_1^\ast=1$, is provided in Figure~\ref{fig:GvO}(b). These results suggest that, with large Grace tuning parameters, Grace substantially outperforms the ridge test in almost all scenarios. The result for the Grace and ridge comparison is similar with $h_n^G / n=1$.

\section{The Grace-Ridge (GraceR) Test}
\label{sec:gracer}

As discussed in Section~\ref{sec:grace}, an informative $\bL$ results in reduced bias of the Grace procedure, by choosing a larger tuning parameter $h$. 
The result in Theorem~\ref{thm:power1} goes beyond just the bias of the Grace procedure. It shows that for certain choices of $\bL$, i.e. when $l$ is close to the true correlation parameter $\rho$, the Grace test can have asymptotically superior power. This additional insight is obtained by accounting for, not just the bias of the Grace procedure, but also its variance, when investigating the power.

However, in practice, there is no guarantee that existing network information truly corresponds to similarities among coefficients, or is complete and accurate. 
To address this issue, we introduce the Grace-ridge (GraceR) test. The estimator used in GraceR incorporates two Grace-type penalties induced by $\bL$ and $\bI$:
\begin{equation}
\label{eq:ridgegrace}
\hat\bbeta(h_G, h_2) = \argmin_\bbeta \left\{\big\|\by-\bX\bbeta\big\|_2^2 + h_G\bbeta^\top\bL\bbeta + h_2\bbeta^\top\bbeta\right\} = \big(n\hat\bSigma+h_G\bL+h_2 \bI\big)^{-1}\bX^\top\by.
\end{equation}
Using data-adaptive choices of tuning parameters $h_G$ and $h_2$, we expect this test to be as powerful as the Grace test if $\bL$ is informative, and as powerful as the GraceI test, otherwise. 

Another advantage of the GraceR over the Grace test is improved bias-variance tradeoff. 
If $\bL$ is (almost) singular, the variance of the Grace test statistic, which depends on the eigenvalues of $(n\hat\bSigma + h\bL)$, could be large even for reasonably large $h$. 
Thus, even though our discussion in Section~\ref{sec:graceest} shows that $(n\hat\bSigma + h\bL)$ is almost surely invertible, with finite samples, its smallest eigenvalue could be very small, if not zero. 
If $\bL$ is informative, $\bL\bbeta$ and hence the bias in \eqref{eq:gracebias} are small. Thus, the rank-deficiency of $(n\hat\bSigma + h\bL)$ can be alleviated by choosing a large value of $h$. However, if $\bL\bbeta$ is non-negligible, choosing a large value of $h$ may result in a large bias, even larger than the ridge estimate. 
to the extent which may offset the benefit from the  variance reduction. 
The finite sample type-I error rate of the Grace test may thus be controlled poorly. By incorporating an additional $\ell_2$ penalty, we can better control the eigenvalues and achieve a better bias-variance trade-off. 

The GraceR optimization problem leads to the following test statistic:
\begin{align}
\hat\bz^{GR} = \hat\bbeta(h_G, h_2) + (n\hat\bSigma+h_G\bL+h_2 \bI)^{-1}(h_G\bL + h_2\bI)\tilde\bbeta.
\end{align}
Similar to Section~\ref{sec:gracetest}, we can write
\begin{align}
\hat z_j^{GR} = \beta_j ^\ast+ Z_j^{GR} +\gamma_j^{GR}, \qquad j = 1,...,p, 
\end{align}
where 
\begin{align*}
Z_j^{GR}|\bX&\sim N \left( 0, n\sigma_\bepsilon^2 \left[ (n\hat\bSigma+h_G\bL+h_2 \bI)^{-1} \hat\bSigma(n\hat\bSigma+h_G\bL+h_2 \bI)^{-1} \right]_{(j,j)} \right), \\
\bgamma^{GR} &\triangleq (n\hat\bSigma+h_G\bL+h_2 \bI)^{-1}(h_G\bL + h_2\bI)(\tilde\bbeta-\bbeta).
\end{align*}
Similar to the Grace test in in Section~\ref{sec:gracetest}, we choose $\tilde\bbeta$ to be an initial lasso estimator, and derive an asymptotic stochastic bound for $\gamma_j^{GR}$ such that $|\gamma_j^{GR}|\precsim^{asy.}\Gamma_j^{GR}$. Equation~\eqref{eq:p} is again used to obtain two-sided $p$-values for $H_0$. 
Theorems~\ref{thm:stbound2} and \ref{thm:suffdet2} parallel the previous results for the Grace test, and establish GraceR's asymptotic control of type-I error rate, and conditions for detection of non-null hypotheses. Proofs of these results are similar to Theorems~\ref{thm:stbound} and \ref{thm:suffdet}, and are hence omitted. 
We first state an alternative to Assumption {\bf{A4}}. This assumption can be justified using an argument similar to that for Assumption {\bf{A4}}, and can also be relaxed with the cost of reduced power for the GraceR test. 
\begin{itemize}
	\item {\bf{A4'}}: $h_G$, $h_2$ and $\bL$ are such that 
	\[
	\left[(n\hat\bSigma+h_G\bL + h_2\bI)^{-1}(h_G\bL+h_2\bI)\right]_{(j,j)} =\mathcal{O}_p\left(\left[\frac{n}{\log{p}}\right]^{\frac{1}{2}-\xi}\right).
	\]
\end{itemize}
\begin{theorem}
\label{thm:stbound2}
Assume Assumptions {\bf{A1}} -- {\bf{A3}} and {\bf{A4'}} are met. The following $\Gamma_j^{GR}$ satisfies the stochastic bound for GraceR.
\begin{align}
\label{eq:gammagr}
\Gamma_j^{GR}\triangleq \left\|\big[(n\hat\bSigma+h_G\bL+h_2 \bI)^{-1}(h_G\bL + h_2\bI)\big]_{(j,-j)}\right\|_\infty\left(\frac{\log{p}}{n}\right)^{\frac{1}{2}-\xi}.
\end{align}
Then, under the null hypothesis, for any $\alpha>0$,
\begin{align}
\limsup_{n\to\infty} Pr\left(\big|\hat\bz_j^{GR}\big|>\alpha\right)\leq\limsup_{n\to\infty}Pr\left(\big|Z_j^{GR}\big|+\Gamma_j^{GR}>\alpha\right).
\end{align}
\end{theorem}
\begin{theorem}
\label{thm:suffdet2}
Assume Assumptions {\bf{A1}} -- {\bf{A3}} and {\bf{A4'}} are met. If for some $h_G > 0$ and $h_2 > 0$, conditional on $\bX$, we have
\begin{align}
\label{eq:tuning2}
\big|\beta_j^\ast\big| >2\Gamma_j^{GR} + q_{(1-\alpha/2)}\sqrt{\Var(Z_j^{GR}|\bX)} + q_{(1-\psi/2)}
\end{align}
for some $0<\alpha<1$ and $0<\psi<1$. Then using the same $h_G$ and $h_2$ in the GraceR test, we get
$
\lim_{n\to\infty} Pr\left(P_j^{GR}\leq\alpha\big|\bX\right) \geq \psi.
$
\end{theorem}

\section{Simulation Experiments}
\label{sec:sim}
In this section, we compare the Grace and GraceR tests with the ridge test \citep{Buhlmann2013} with small tuning parameters, low-dimensional projection estimator (LDPE) for inference \citep{ZhangZhang2014,vandeGeeretal2014} and the GraceI test. 
To this end, we consider a graph similar to \citet{LiLi2008}, with 50 hub covariates (genes), each connected to 9 other satellite covariates (genes). The 9 satellite covariates are not connected with each other, nor are covariates in different hub-satellite clusters. In total the graph includes  $p=500$ covariates and 450 edges; see Figure {S1} in Section~\ref{sec:supp} for an illustration with 5 hub-satellite clusters.
We build the underlying true Laplacian matrix $\bL^*$ according to the graph with all edge weights equal 1. 

To assess the effect of inaccurate or incomplete network information, we also consider variants of the Grace and GraceR tests with incorrectly specified graphs, where a number of randomly selected edges are added or removed. 
The number of removed or added (perturbed) edges relative to the true graph is $\textrm{NPE}\in\{$\textsc{-165, -70, -10, 0, 15, 135, 350}$\}$, with negative and positive numbers indicating removals and additions of edges, respectively. For example, $\textrm{NPE}$=\textsc{-165} indicates 165 of the 450 edges in the true graph represented by $\bL^*$ are randomly removed in the perturbed graph with corresponding perturbed Laplacian matrix $\bL$. This represents the case with incomplete network information. On the other hands, $\textrm{NPE} = 350$ indicates that in addition to the 450 true edges in $\bL^*$, we also randomly add 350 wrong edges to $\bL$. 
The $\textrm{NPE}$ values considered correspond to similar normalized spectral differences for settings where edges are removed or added, i.e. $\|\bL-\bL^*\|_2/\|\bL^*\|_2\approx (0.75, 0.50, 0.25, 0, 0.25, 0.50, 0.75)$. Thus, the size of perturbation to the graph is roughly the same with $\textrm{NPE}=\textsc{-}165$ and 350. The perturbed penalty weight matrix $\bL$ is then used in the Grace and GraceR tests.  
Since $(\bX^\top\bX+h\bL)$ may not be invertible, for Grace, we add a value of 0.01 to the diagonal entries of $\bL$ to make it positive definite. No such correction is needed for GraceR and GraceI because of the $\ell_2$ penalty.

In each simulation replicate, we generate $n = 100$ independent samples, where for the 50 hub covariates in each sample, $x_k^{hub}\sim^{iid}N(0, 1)$, $k = 1,..., 50$, and for the 9 satellite covariates in the $k$-th hub-satellite cluster, $x_l^{hub_k}\sim^{iid}N(0.9\times x_k^{hub}, 0.9)$, $l=1,...,9$, $k = 1,..., 50$. This is equivalent to  simulating $\bx^i\sim^{iid} N_p(\bzero, \bSigma)$ for $i =1,...,100$ with $\bSigma = (\bL^* + 0.11\times\bI)^{-1}$, where $\bL^*$ corresponds to the partial covariance structure of the covariates.

We consider a sparse model in which covariates in the first hub-satellite cluster are equally associated with the outcome, and those in the other 49 clusters are not. Specifically,  we let
\[
\bbeta^\ast\triangleq\frac{1}{\sqrt{10}}(\underbrace{1,...,1}_{10}, \underbrace{0,...,0}_{p-10})^\top. 
\] 
We then simulate 
$
\by=\bX\bbeta^\ast+\bepsilon,
$
with $\bepsilon\sim N_n(\bzero, \sigma^2_{\epsilon}\bI_n)$, and consider $\sigma_\bepsilon\in\{9.5, 6.3, 4.8\}$ to produce expected $R^2=1-\sigma_\bepsilon^2/\Var(\by)\in\{0.1, 0.2, 0.3\}$.

\begin{figure}
\caption{Comparison of powers and type-I error rates of different testing methods, along with their 95\% confidence bands. Testing methods include LDPE  \citep{ZhangZhang2014, vandeGeeretal2014}, ridge \citep{Buhlmann2013}, GraceI, Grace and GraceR tests. Filled circles ($\bullet$) corresponds to powers, whereas crosses ($\times$) are type-I error rates. Numbers on $x$-axis for Grace and GraceR tests refer to the number of perturbed edges (\textrm{NPE}) in the network used for testing, compared to the true network used to generate the data.} 
\label{fig:power}
\centering
\includegraphics[angle = 90, height = 20cm]{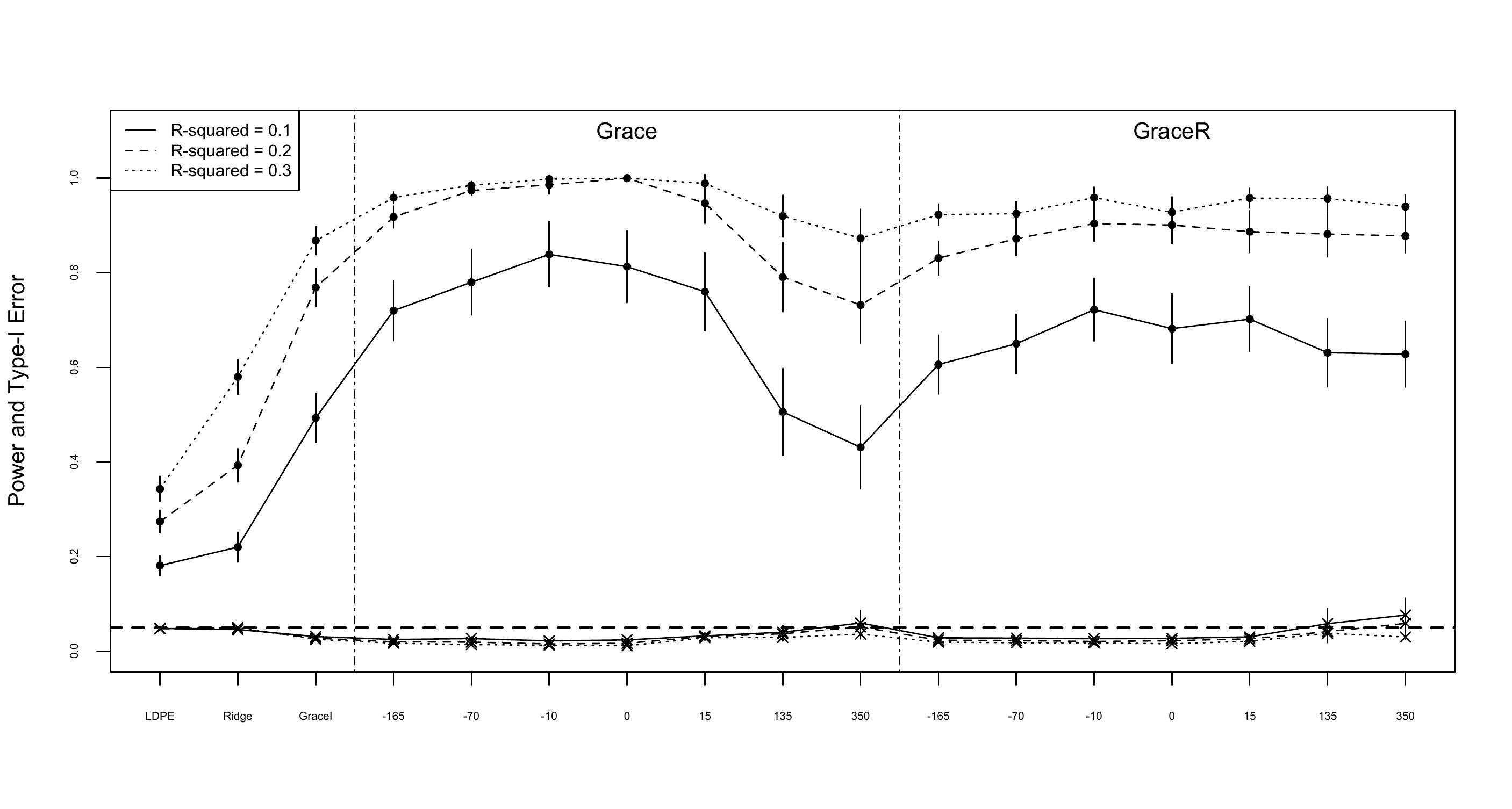} 
\end{figure}

Throughout the simulation iterations, $\bL^*$ and $\bbeta^\ast$ are kept fixed, and $\bL$, $\bX$ and $\bepsilon$ are randomly generated in each repetition. We set the sparsity parameter $\xi = 0.05$, and $h_{Lasso}=4\hat\sigma_\bepsilon\sqrt{3\log{p}/n}$, where $\hat\sigma_\bepsilon$ is calculated using the scaled lasso \citep{SunZhang2012}. 
As suggested in \citet{Buhlmann2013}, the tuning parameter for the ridge test is set to 1. 
Tuning parameters for LDPE, Grace, GraceR and GraceI are chosen by 10-fold CV.
We use two-sided significance level $\alpha=0.05$ and calculate the average and  standard error of powers from 10 non-zero coefficients and the type-I error rates of each test from 490 zero coefficients. 
Figure~\ref{fig:power} summarizes the mean powers and type-I error rates of tests across $B=100$ simulated data sets, along with the corresponding 95\% confidence intervals. Detail values of powers and type-I error rates, as well as an expanded simulation with a larger range of NPE, are available in Section~\ref{sec:supp}.

Comparing the power of the tests, it can be seen that the Grace test with correct choices of $\bL$ ($\textrm{NPE} = 0$) results in highest power. The  performance of the Grace test, however, deteriorates as $\bL$ becomes less accurate. The performance of the GraceR test is, on the other hand, more stable. It is close to the Grace test when the observed $\bL$ is close to the truth, and is roughly as good as the GraceI test when $\bL$ is significantly inaccurate. As expected, our testing procedures asymptotically control the type-I error rate, in that observed type-I error rates are not significantly different from $\alpha=0.05$. 
%

\section{Analysis of TCGA Prostate Cancer Data}
\label{sec:rd}
We examine the Grace and GraceR tests on a prostate adenocarcinoma dataset from The Cancer Genome Atlas (TCGA) collected from prostate tumor biopsies. After removing samples with missing measurements, we obtain a dataset with $n=321$ samples. For each sample, the prostate-specific antigen (PSA) level and the RNA sequences of 4739 genes are available. Genetic network information for these genes is obtained from the Kyoto Encyclopedia of Genes and Genomes (KEGG), resulting in a dataset with $p=3450$ genes and $|E|=38541$ edges.

We center the outcome and center and scale the covariates. For the Grace and GraceR tests, we set the sparsity parameter $\xi=0.05$ and $h_{Lasso}=4\hat\sigma_\bepsilon\sqrt{3\log{p}/n}$, where $\hat\sigma_\bepsilon$ is calculated using the scaled lasso \citep{SunZhang2012}.
We control the false discovery rate at $\alpha = 0.05$ level using the method of \citet{BenjaminiYekutieli2001}. 

To increase the chance of selecting ``hub'' genes, we use the normalized Laplacian matrix $\bL^{(norm)} = \mathbf{D}^{-1/2} \bL \mathbf{D}^{-1/2}$, where $\mathbf{D}$ is the diagonal degree matrix for the KEGG network with edge weights set to 1.
The Grace penalty induced by the normalized Laplacian matrix encourages smoothness of coefficient estimates based on the degrees of respective nodes, $\bbeta^\top\bL^{(norm)}\bbeta = \sum_{u\sim v} (\bbeta_u/\sqrt{d_u} - \bbeta_v/\sqrt{d_v})^2 w(u,v)$ \citep{LiLi2008}. We add 0.001 to the diagonal entries of $\bL^{(norm)}$ to induce positive definitiveness in the Grace test.

\begin{figure}[h]
\centering
\caption{Results of analysis of TCGA prostate cancer data using the (a) \emph{Grace} and (b) \emph{GraceR} tests after adjusting for FDR at 0.05 level. In each case, genes found to be significantly associated with PSA level are shown, along with their interactions based on information from KEGG.} \label{fig:RD}
\includegraphics[width = 15cm]{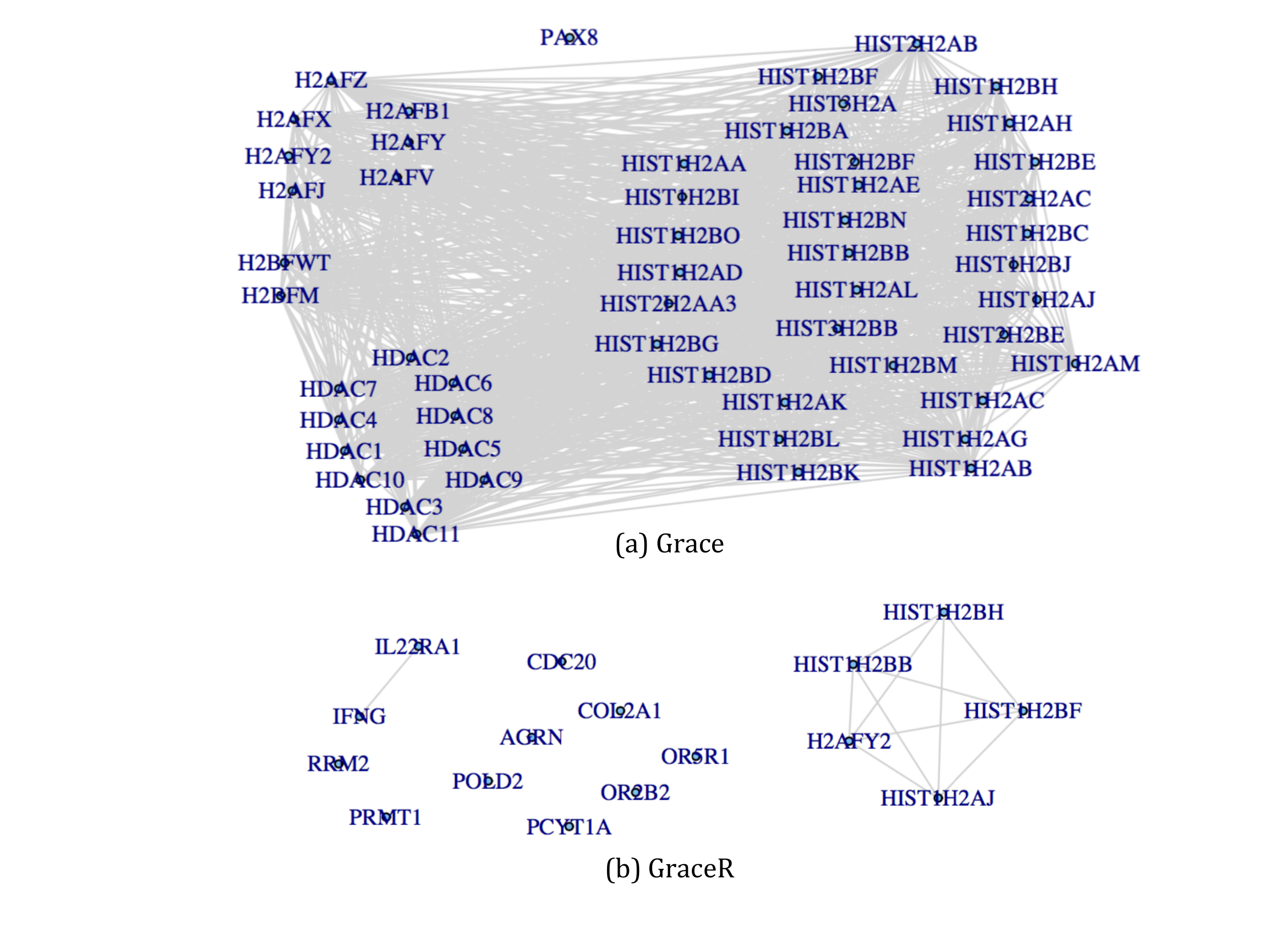}
\end{figure}

As shown in Figure~\ref{fig:RD}(a), the Grace test with tuning parameter 
selected by 10-fold CV identifies 54 genes that are associated with PSA level. They consist of 42 histone genes, 11 histone deacetylase (HDAC) genes and the paired box gene 8 (PAX8). Histone and HDAC genes are densely connected in the KEGG network. With the network smoothing penalty, the Grace regression coefficients of histone and HDAC genes are all positive with a similar magnitude. Existing literature indicates that the histone and HDAC genes are associated with the occurrence, progression, clinical outcomes or recurrence of prostate cancer.
Figure~\ref{fig:RD}(b) shows the result for the GraceR test. 
GraceR identifies 5 histone genes, which are also identified by the Grace test. In addition, GraceR identifies 11 genes that are not identified by Grace. Prior work has identified 9 of those 11 genes to be associated with PSA level or the severity and stage of cancer. Additional details about existing evidence in support of genes identified using Grace and GraceR tests, as well as extended results on  prediction performance and stability of the Grace test are provided in Section~\ref{sec:supp}. 
%

As a comparison, the GraceI test with 10-fold CV identifies 16 disconnected genes, 11 of them are also identified by the GraceR test. Ridge test \citep{Buhlmann2013} with tuning parameter $h_2 = 1$ identifies 4 disconnected genes, which are also identified by the GraceR test. The low-dimensional projection estimator (LDPE) with tuning parameters chosen by 10-fold CV identifies 10 disconnected genes. Seven of these genes are identified by GraceR and two by Grace.

\section{Discussion}\label{sec:disc}
In this paper, we proposed the Grace and GraceR tests that incorporate external graphical information regarding the similarity between covariates. Such external information is presented in the form of a penalty weight matrix $\bL$, which is considered to be the (normalized) graph Laplacian matrix in this paper. However, any positive semi-definite matrix can be used as $\bL$. The proposed inference framework thus allows researchers in different fields to incorporate relevant external information through $\bL$. For example, we can use various distance and kernel metrics that measure the (dis)similarity between species in phylogenetic studies. We can also use the adaptive graph Laplacian matrix \citep{LiLi2010} so that coefficients of negatively correlated covariates are penalized to have the opposite signs. Regardless of the choice of $\bL$, our proposed procedures asymptotically control the type-I error rate; the power of the Grace test, however, depends on the informativeness of $\bL$. The power of the GraceR test is on the other hand less dependent on the choice of $\bL$.

The Grace test introduced in this paper is not scale invariant. That is, the Grace test with the same tuning parameter could produce different $p$-values with data $(\bX,\by)$ and $(\bX, k\by)$, where $k\neq 1$ is a constant. This is clear as the test statistic $\hat z_j$ depends on $\by$ whereas the stochastic bound $\Gamma_j^G$ does not. To make the Grace and GraceR tests scale invariant, we can simply choose the tuning parameter for our lasso initial estimator to be $h_{Lasso}=C\sigma_\bepsilon\sqrt{\log{p}/n}$ with a constant $C>2\sqrt{2}$. \citet{SunZhang2012} show that the lasso is scale invariant in this case. We would also need to use scaled invariant stochastic bounds $\tilde\Gamma_j^G\triangleq \sigma_\bepsilon\Gamma_j^G$ and $\tilde\Gamma_j^{GR}\triangleq \sigma_\bepsilon\Gamma_j^{GR}$ in our Grace and GraceR tests. Note that multiplying any constant in $\Gamma_j^G$ and $\Gamma_j^{GR}$does not change our asymptotic control of the type-I error rate. 

In this paper, cross validation (CV) is used to choose tuning parameters of the Grace and GraceR tests. However, CV does not directly maximize the power of these tests. 
Selection of tuning parameters for optimal testing performance can be a fruitful direction of future research. 
Another useful extension of the proposed framework is its adaptation to generalized linear models (GLM).

\section*{Acknowledgements}
We would like to thank Dr. Ruben Dezeure and Dr. Peter B\"{u}hlmann of the Seminar for Statistics of the Department of Mathematics at ETH Z\"{u}rich for providing the code for LDPE.

\vspace*{-8pt}

\section{Supplementary Materials}
\label{sec:supp}

\subsection{Proof of Lemma \ref{lem:bias}}

\begin{proof} Given that $(n\hat\bSigma+h\bL)$ is invertible and $h>0$, we have
\begin{align*}
\mathbf{Bias}\big(\hat\bbeta(h)\big|\bX\big) &=\E\big(\hat{\bbeta}(h)\big|\bX\big) -\bbeta^\ast \\
& = (n\hat\bSigma+h\bL)^{-1}n\hat\bSigma\bbeta^\ast - (n\hat\bSigma+h\bL)^{-1}(n\hat\bSigma+h\bL)\bbeta^\ast\\
&=-(n\hat\bSigma+h\bL)^{-1}h\bL\bbeta^\ast,
\end{align*}
which is equal to $\bzero$ if and only if
$
	\bL\bbeta^\ast =\bzero.
$
We know that
\[
(n\hat\bSigma+h\bL)^{-1}\preceq \frac{1}{{\lambda_{0}(n\hat\bSigma+h\bL)}}\bI.
\]
Therefore,
\begin{align*}
\big\|\mathbf{Bias}(\hat{\bbeta}(h))\big|\bX\big\|_2 &= h\sqrt{(\bL\bbeta^\ast)^\top(n\hat\bSigma+h\bL)^{-2}(\bL\bbeta^\ast)} \\
&\leq h\sqrt{(\bL\bbeta^\ast)^\top\frac{1}{{\lambda_{0}(n\hat\bSigma+h\bL)}^2}(\bL\bbeta^\ast)} \\
&= \frac{h\|\bL\bbeta^\ast\|_2}{\lambda_{0}(n\hat\bSigma+h\bL)}.
\end{align*}
\end{proof}

\subsection{Proof of Theorem \ref{thm:stbound}}

\begin{proof}
Under the null hypothesis $H_0: \beta_j^\ast=0$, we have
\begin{align*}
\big|\gamma_j^{G}\big| &=h\big|(n\hat\bSigma+h\bL)^{-1}\bL(\tilde\bbeta-\bbeta^\ast)\big|_j \\
&=h\big|\sum_{i=1}^p\big[(n\hat\bSigma+h\bL)^{-1}\bL\big]_{(j,i)}(\tilde\beta_i-\beta^\ast_i)\big| \\
&\leq h\big|\sum_{i:i\neq j}\big[(n\hat\bSigma+h\bL)^{-1}\bL\big]_{(j,i)}(\tilde\beta_i-\beta^\ast_i)\big| + h\big|\big[(n\hat\bSigma+h\bL)^{-1}\bL\big]_{(j,j)}\tilde\beta_j\big|\\
&\leq h\big\|\big[(n\hat\bSigma+h\bL)^{-1}\bL\big]_{(j,-j)}\big\|_\infty \big\|\tilde\bbeta-\bbeta^\ast\big\|_1 + h\big|\big[(n\hat\bSigma+h\bL)^{-1}\bL\big]_{(j,j)}\tilde\beta_j\big|
\end{align*}

Based on \citet{BuhlmannvandeGeer2011}, Chapter 6.12, with Gaussian design, if the $\bSigma$-compatibility condition is met for the set $S_0$ with compatibility constant $\phi_{\bSigma}$, with probability tending to 1, the condition is also met for $\hat\bSigma$ with compatibility constant $\phi_{\hat\bSigma} > \phi_{\bSigma} / 2$. 
Moroever,
with $h_{Lasso}\asymp \sqrt{\log{p}/n}$ and the  $\hat\bSigma$-compatibility condition for the set $S_0$, with probability tending to 1, we have
\[
\big\|\tilde\bbeta-\bbeta^\ast\big\|_1\leq 4\frac{h_{Lasso} s_0}{\phi_{\hat\bSigma}^2}.
\]
Then, because $s_0 = \smallO([n/\log{p}]^\xi)$ and $\liminf\phi_{\hat\bSigma}^2>d/2>0$, we get
\[
\big\|\tilde\bbeta-\bbeta^\ast\big\|_1=\small\smallO_p\left(\big(\frac{\log{p}}{n}\big)^{\frac{1}{2}-\xi}\right).
\]
On the other hand, by Assumption A4, $\big((n\hat\bSigma+h\bL)^{-1}h\bL\big)_{(j,j)} =\mathcal{O}_p\big((n/\log{p})^{1/2-\xi}\big)$. Thus
\[
h\big|\big[(n\hat\bSigma+h\bL)^{-1}\bL\big]_{(j,j)}\tilde\beta_j\big| = \big|\big[(n\hat\bSigma+h\bL)^{-1}h\bL\big]_{(j,j)}\big|\big|\tilde\beta_j - \beta^\ast_j\big|=\small\smallO_p(1),
\]
and hence
\[
Pr\left(\big|\gamma_j^{G}\big| \leq h\big\|[(n\hat\bSigma+h\bL)^{-1}\bL]_{(j,-j)}\big\|_\infty\big(\frac{\log{p}}{n}\big)^{\frac{1}{2}-\xi}\right)\to 1,
\]
where the right hand side is $\Gamma^G_j$. We can thus write
\begin{align*}
\big|\hat\bz_j^G\big| &= \big|Z_j^G +\gamma_j^{G}\big| \\
& \leq \big|Z_j^G\big| +|\gamma_j^{G}\big|  \\
&\precsim^{asy.}\big|Z_j^G\big|+\Gamma_j^G.
\end{align*}
\end{proof}

\subsection{Proof of Theorem \ref{thm:suffdet}}

\begin{proof}
Given \eqref{eq:p}, conditional on $\bX$,
the objective of $P_j^G \leq \alpha$ is satisfied if $\big|\hat\bz_j^G\big| \geq\Gamma_j^G + q_{(1-\alpha/2)}\sqrt{\Var(Z_j^G|\bX)}$. According to Equation (\ref{eq:keyrel}), this is equivalent of $\big|\beta^\ast_j+Z_j^G + \gamma_j^G\big| \geq\Gamma_j^G + q_{(1-\alpha/2)}\sqrt{\Var(Z_j^G|\bX)}$, which is satisfied if
\[
\big|\beta^\ast_j\big| - \big|\gamma_j^G\big|-\big|Z_j^G\big| \geq\Gamma_j^G + q_{(1-\alpha/2)}\sqrt{\Var(Z_j^G|\bX)}.
\]
This holds with probability at least $\psi$ if 
\[
\big|\beta^\ast_j\big| - \big|\gamma_j^G\big| \geq\Gamma_j^G + q_{(1-\alpha/2)}\sqrt{\Var(Z_j^G|\bX)} + q_{(1-\psi/2)}.
\]

We know that with probability tending to 1, $\big|\gamma_j^G\big|\leq \Gamma_j^G$.  Therefore, conditional on $\bX$, we have $P_j^G \leq \alpha_L$ with probability tending to at least $\psi$,  if
\[
\big|\beta^\ast_j\big| >2\Gamma_j^G + q_{(1-\alpha/2)}\sqrt{\Var(Z_j^G|\bX)} + q_{(1-\psi/2)}.
\]

\end{proof}

\subsection{Proof of Theorem \ref{thm:power1}}

\begin{proof}
\textbf{a)} We note that $P_1^G/P_1^{GI}\leq 1$ is equivalent of  
\[
\frac{\left(\big|\hat\bz_1^{GI}\big|-\Gamma_1^{GI}\right)_+/\sqrt{\Var(Z_1^{GI}|\bX)}}{\left(\big|\hat\bz_1^G\big|-\Gamma_1^G\right)_+/\sqrt{\Var(Z_1^G|\bX)}}\leq1. 
\]
We first write out those components for the Grace test:
\begin{align*}
\hat\bz_1^G&=\big((\bX^\top\bX+h_n^G \bL)^{-1}(\bX^\top\by+h_n^G\bL\tilde\bbeta)\big)_1 \\
&= \frac{(n+h_n^G)\bx_1^\top\by-(n\rho+h_n^Gl)\bx_2^\top\by + h_n^G\tilde\beta_1(n+h_n^G-n\rho l-h_n^Gl^2)+nh_n^G\tilde\beta_2(l-\rho)}{(n+h_n^G)^2-(n\rho+h_n^G l)^2}; \\
\Gamma_1^G & = \left|h_n^G\big[(\bX^\top\bX+h_n^G\bL)^{-1}\bL\big]_{(1, -1)}\right|\left(\frac{\log{p}}{n}\right)^{\frac{1}{2}-\xi} \\
&= \left|h_n^G\big[(\bX^\top\bX+h_n^G\bL)^{-1}\bL\big]_{(1, 2)}\right|\left(\frac{\log{p}}{n}\right)^{\frac{1}{2}-\xi}\\
&= \frac{|nh_n^Gl-nh_n^G\rho|}{(n+h_n^G)^2-(n\rho+h_n^G l)^2}\left(\frac{\log{p}}{n}\right)^{\frac{1}{2}-\xi}; \\
\Var(Z_1^G|\bX)& = \sigma^2_\bepsilon\left[(\bX^\top\bX+h_n^G \bL)^{-1}\bX^\top\bX(\bX^\top\bX+h_n^G \bL)^{-1}\right]_{(1, 1)} \\
& =  \sigma^2_\bepsilon\frac{(n^3+2h_n^G n^2)(1-\rho^2)+n(h_n^G)^2(1+l^2-2l\rho)}{[(n+h_n^G)^2-(n\rho+h_n^G l)^2]^2}.
\end{align*}
We can also write out those components for the GraceI test likewise with $l=0$.

In the proof of Theorem~\ref{thm:stbound}, we have shown that $Pr\left(\big\|\tilde\bbeta-\bbeta^\ast\big\|_1\leq 4h_{Lasso}s_0/\phi^2_{\hat\bSigma}\right)\to1$. With $h_{Lasso}=\mathcal{O}(\log{p}/n)$, $s_0=\smallO([n/\log{p}]^\xi)$ for some $0\leq\xi<1/2$, $\liminf\phi_{\hat\bSigma}>d/2>0$, and $p=\mathcal{O}(\exp(n^\nu))$ for some $0\leq\nu<1$, we have $\|\tilde\bbeta-\bbeta\|_1=\small\smallO_p(1)$. Thus we get 
\[
\tilde\beta_1 = \beta_1^\ast+\small\smallO_p(1), \hspace{0.5in} \tilde\beta_2 = \beta_2^\ast+\small\smallO_p(1). 
\]
We also note that since our design matrix is scaled, we get
\begin{align*}
\bx_1^\top\by&=\bx_1^\top\bx_1\beta^\ast_1 + \bx_1^\top\bx_2\beta^\ast_2 + \bx_1^\top\bepsilon = n\beta^\ast_1+n\rho\beta^\ast_2 + nE, \\
\bx_2^\top\by&=\bx_2^\top\bx_1\beta^\ast_1 + \bx_2^\top\bx_2\beta^\ast_2 + \bx_2^\top\bepsilon = n\rho\beta^\ast_1+n\beta^\ast_2 + nE, 
\end{align*}
where $E\sim N\left(\bzero, \sigma^2_\bepsilon/n\right)=\small\smallO_p(1)$. 

Define $k_n^G\triangleq h_n^G/n$ and $k_n^{GI}\triangleq h_n^{GI}/n$. With some algebra, We get
\begin{align}
\label{ts:G}
\frac{\left(|\hat\bz_1^G|-\Gamma_1^G\right)_+}{\sqrt{\Var(Z_1^G|\bX)}} = \frac{\sqrt{n}\left[|(k_n^G+1)^2-(\rho+lk_n^G)^2 + \smallO_p(1)|\cdot|\beta_1^\ast| - (\log{p}/n)^{1/2-\xi}\cdot|k_n^G(l-\rho)|\right]_+}{\sigma_\bepsilon\sqrt{(1+2k_n^G)(1-\rho^2) + (k_n^G)^2(1+l^2-2l\rho)}}.
\end{align}
Similarly for the GraceI, we get
\begin{align}
\label{ts:GI}
\frac{\left(|\hat\bz_1^{GI}|-\Gamma_1^{GI}\right)_+}{\sqrt{\Var(Z_1^{GI}|\bX)}} = \frac{\sqrt{n}\left[|(k_n^{GI}+1)^2-\rho^2 + \smallO_p(1)|\cdot|\beta_1^\ast| - (\log{p}/n)^{1/2-\xi}\cdot|k_n^{GI}\rho|\right]_+}{\sigma_\bepsilon\sqrt{(1+2k_n^{GI})(1-\rho^2) + (k_n^{GI})^2}}.
\end{align}

We observe that $k_n^{GI}+1>1\geq |\rho|$ and $k_n^G+1\geq|l|k_n^G+|\rho|\geq |\rho+lk_n^G|$. We plug in those two inequalities into Equation ~\eqref{ts:G} and ~\eqref{ts:GI}. Hence, conditional on the design matrix $\bX$, $P_1^G/P_1^{GI}\leq 1$ with probability tending to 1 if
\begin{align*}
&\lim_{n\to\infty}\frac{\left\{\big[(k_n^G+1)^2-(\rho+lk_n^G)^2\big]\cdot|\beta_1^\ast| - (\log{p}/n)^{1/2-\xi}\cdot|k_n^G(l-\rho)|\right\}_+}{\sqrt{(1+2k_n^G)(1-\rho^2) + (k_n^G)^2(1+l^2-2l\rho)}} \\
\geq &\lim_{n\to\infty}\frac{\left\{\big[(k_n^{GI}+1)^2-\rho^2\big]\cdot|\beta_1^\ast| - (\log{p}/n)^{1/2-\xi}\cdot|k_n^{GI}\rho|\right\}_+}{\sqrt{(1+2k_n^{GI})(1-\rho^2) + (k_n^{GI})^2}}.
\end{align*}

Note that for any two real numbers $f$ and $g$, $f\geq g$ implies $f_+\geq g_+$. Thus, conditional on the design matrix $\bX$, $P_1^G/P_1^{GI}\leq 1$ with probability tending to 1 if
\begin{align}
\label{ts:GGI}
&\lim_{n\to\infty}\frac{\big[(k_n^G+1)^2-(\rho+lk_n^G)^2\big]\cdot|\beta_1^\ast| - (\log{p}/n)^{1/2-\xi}\cdot|k_n^G(l-\rho)|}{\sqrt{(1+2k_n^G)(1-\rho^2) + (k_n^G)^2(1+l^2-2l\rho)}} \nonumber\\
\geq &\lim_{n\to\infty}\frac{\big[(k_n^{GI}+1)^2-\rho^2\big]\cdot|\beta_1^\ast| - (\log{p}/n)^{1/2-\xi}\cdot|k_n^{GI}\rho|}{\sqrt{(1+2k_n^{GI})(1-\rho^2) + (k_n^{GI})^2}}.
\end{align}

If we assume $k_n^G=k_n^{GI}=k\to\infty$, Inequality~\eqref{ts:GGI} is satisfied if 
\begin{align}
&\lim_{n\to\infty}\frac{\big[(k+1)^2-(\rho+lk)^2\big]\cdot|\beta_1^\ast| - (\log{p}/n)^{1/2-\xi}\cdot|k(l-\rho)|}{\big[(k+1)^2-\rho^2\big]\cdot|\beta_1^\ast| - (\log{p}/n)^{1/2-\xi}\cdot|k\rho|} \nonumber \\
\times&\frac{\sqrt{(1+2k)(1-\rho^2) + k^2}}{\sqrt{(1+2k)(1-\rho^2) + k^2(1+l^2-2l\rho)}} \nonumber \\ 
=&\lim_{n\to\infty}\frac{\big[(1-l^2)+(2-2l\rho)/k+(1-\rho^2)/k^2\big]\cdot|\beta_1^\ast| - (\log{p}/n)^{1/2-\xi}\cdot|(l-\rho)/k|}{\big[1+2/k+(1-\rho^2)/k^2\big]\cdot|\beta_1^\ast| - (\log{p}/n)^{1/2-\xi}\cdot|\rho/k|} \nonumber\\
\times& \frac{\sqrt{1 + (2-2\rho^2)/k + (1-\rho^2)/k^2 }}{\sqrt{(1+l^2-2l\rho) + (2-2\rho^2)/k + (1-\rho^2)/k^2}}  \nonumber\\
=&\frac{(1-l^2)}{\sqrt{(1+l^2-2l\rho)}} \geq1.
\end{align}
The last equality holds because $p=\mathcal{O}(\exp(n^\nu))$ for some $0\leq\nu<1$ implies that $\log{p}/n\to0$.

For the ridge test, we assume $h_n^R=\mathcal{O}(1)$. Thus with some algebra we can similarly write out the ridge test objective:
\begin{align}
\label{ts:R}
\frac{|\hat\bz_1^R|}{\sqrt{\Var(Z_1^R|\bX)}} = \frac{\sqrt{n}|1-\rho^2 + \smallO_p(1)|\cdot|\beta_1^\ast|}{\sigma_\bepsilon\sqrt{(1-\rho^2)+\smallO(1)}}.
\end{align}

\textbf{b)} Thus, conditional on $\bX$, we get $P_1^G/P_1^{R}\leq 1$ with probability tending to 1 if
\begin{align}
\label{ts:GR}
\lim_{n\to\infty}\frac{\big((k_n^G+1)^2-(\rho+lk_n^G)^2\big)\cdot|\beta_1^\ast| - (\log{p}/n)^{1/2-\xi}\cdot|k_n^G(l-\rho)|}{\sqrt{(1+2k_n^G)(1-\rho^2) + (k_n^G)^2(1+l^2-2l\rho)}}\geq \sqrt{1-\rho^2}\cdot|\beta_1^\ast|.
\end{align}

\textbf{c)} We also have$P_1^{GI}/P_1^{R}\leq 1$ with probability tending to 1 if
\begin{align}
\label{ts:GIR}
\lim_{n\to\infty}\frac{\big((k_n^{GI}+1)^2-\rho^2\big)\cdot|\beta_1^\ast| - (\log{p}/n)^{1/2-\xi}\cdot|k_n^{GI}\rho|}{\sqrt{(1+2k_n^{GI})(1-\rho^2) + (k_n^{GI})^2}}\geq \sqrt{1-\rho^2}\cdot|\beta_1^\ast|.
\end{align}

\end{proof}

\subsection{Illustration of the Graph Structure in the Simulation Study}
\begin{figure}[h]
\caption{An illustration of the graph structure with 5 hub-satellite clusters.} 
\label{fig:graph}
\centering
\includegraphics[width = 12cm]{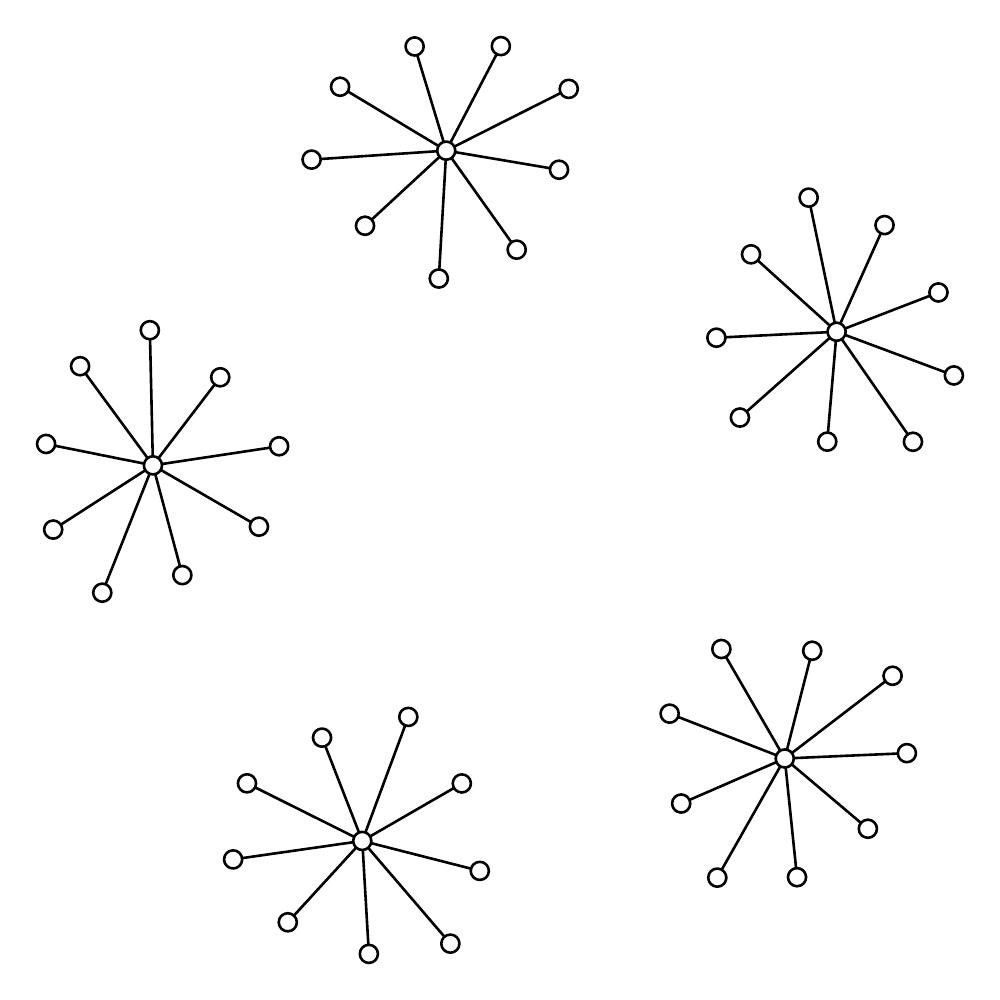} 
\end{figure}

Figure~\ref{fig:graph} shows the graph structure used in the simulation study with 5 hub-satellite clusters. In the simulation study, we use 50 such hub-satellite clusters.

\subsection{Additional Details for Analysis of TCGA Data}
\subsubsection{Biological Evidence}
In this section, we summarize some of the biological evidences in support of the association between genes identified by the Grace and GraceR tests with the onset, progression and severity of prostate cancer, as well as PSA level. 

As pointed out in the main paper, the Grace and GraceR tests identify a number of histone genes and histone deacetylase (HDAC) genes. Previous research indicates that histone genes are associated with the occurrence, clinical outcomes and recurrence of prostate cancer \citep{Seligsonetal2005, Keetal2009}. The pathological role of HDAC genes on the onset and progression of prostate cancer have also been previously studied \citep{Halkidouetal2004, Chenetal2007, AbbasGupta2008}. 

In addition to the highly connected histone and HDAC genes, the GraceR test also identifies some disconnected genes. Prior works shows that the expression of ribonucleoside-diphosphate reductase subunit M2 (RRM2) is associated with higher Gleason scores, which correlate with the severity of prostate cancer \citep{Huangetal2014}. Protein arginine methyltransferase 1 (PRMT1) may also have an effect on the proliferation of prostate cancer cells \citep{Yuetal2009}. Activation of olfactory receptors (OR) prevents proliferation of prostate cancer cells \citep{Neuhausetal2009}. Interferon-$\gamma$ (IFNG) plays a role in the differentiation of human prostate basal-epithelial cells \citep{Untergasseretal2005}. IFNG is connected to the interleukin receptor 22 $\alpha1$ (IL22RA1), the role of which related to prostate cancer is unknown. However, several earlier studies point out the associations between prostate cancer and several other interleukin receptors in the Janus kinase and signal transducer and activator of transcription (JAK-STAT) activating family, including IL 6, 8, 11, 13 and 17 genes\citep{Culigetal2005, Inoueetal2000, Campbelletal2001, Mainietal1997, Zhangetal2012}. Cell-division cycle genes (CDC) may also be associated with various cancers. The association between collagen type 2 $\alpha1$ (COL2A1) and prostate cancer is also not known, but other collagen genes, including type 1 $\alpha2\beta1$, type 4 $\alpha5$ and $\alpha6$, have been shown to be associated with prostate cancer progression \citep{Halletal2008, Dehanetal1997}.  Although the association between phosphate cytidylyltransferase 1 choline-$\alpha$ (PCYT1A) and prostate cancer or PSA level is not known, \citet{Vaezietal2014} shows that PCYT1A is a prognostic factor in survival for patients with lung and head and neck squamous cell carcinomas.

\subsubsection{Stability of the Grace Test to the Tuning Parameter}
\begin{figure}[h]
\caption{Number of genes identified by the Grace test in the TCGA data against the tuning parameter of the Grace test, $h_G$. The red dashed line corresponds to the choice made by 10-fold CV ($h_G=\exp(14.2)$).} 
\label{fig:ngenes}
\centering
\includegraphics[width = 12cm]{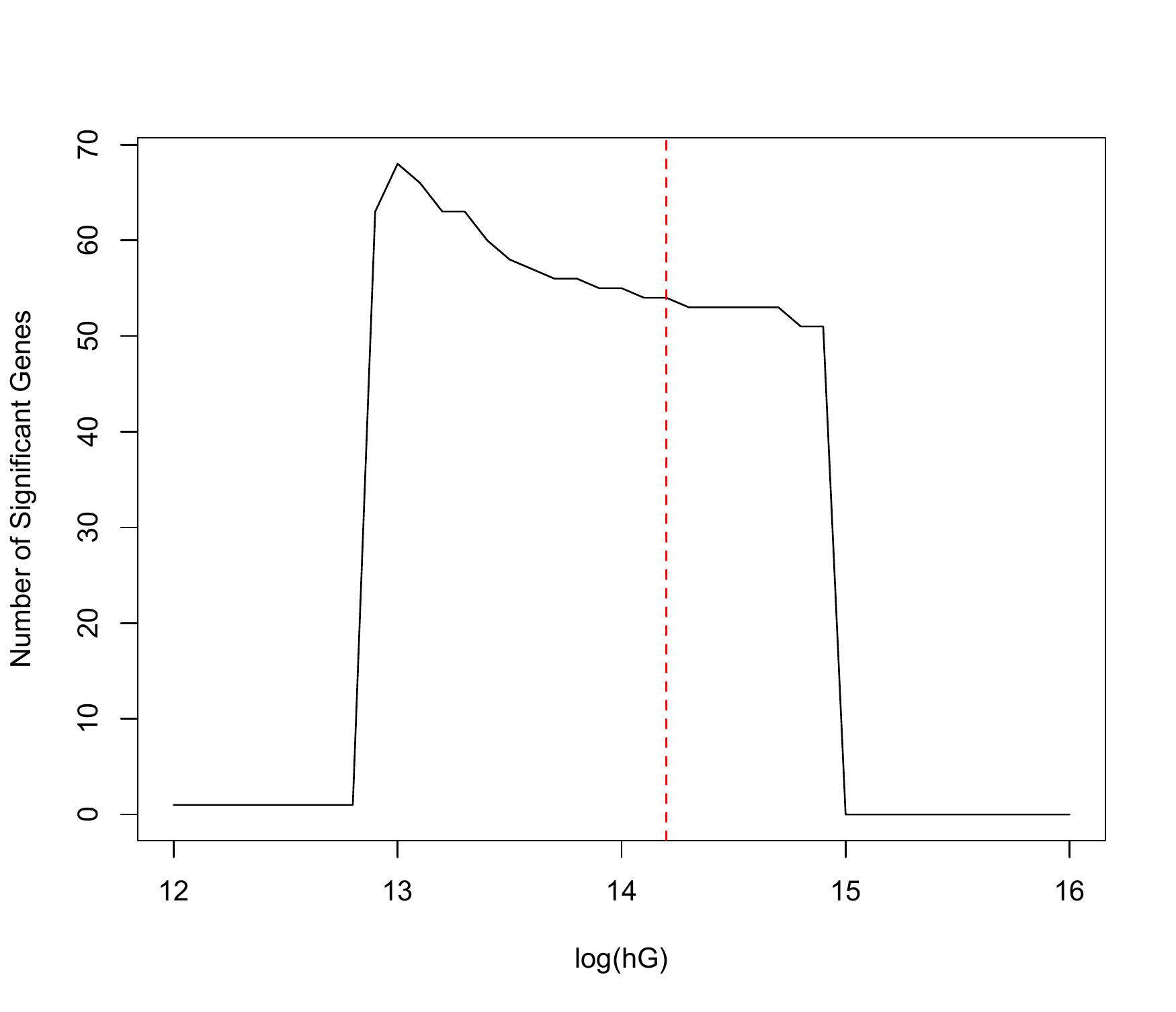} 
\end{figure}

Figure~\ref{fig:ngenes} shows the number of significant genes identified by the Grace test in the TCGA data against various values of $h_G$.  The results indicate that the number of genes found by the Grace test is relatively stable for a range of tuning parameters including the CV choice. On the other hand, very few genes are identified when the tuning parameter is too small or too large. This is because, with small tuning parameters, the variance is large and thus no gene is statistically significant. On the other hand, with large tuning parameters, the stochastic bound $\Gamma_j$ dominates $\hat z_j$. Note that above results of power do not contradict Theorem~\ref{thm:power1}, which shows the \textit{asymptotic} power of the Grace test improves as we use larger $h_G$. A vital condition for Theorem~\ref{thm:power1} to hold is $\|\tilde\bbeta-\bbeta\|_1=\smallO_p(1)$. 

\subsubsection{Stability of the Grace Test to the Network}
\begin{figure}[h]
\caption{Number of genes that are significant using both the KEGG network and the perturbed network against the number of perturbed edges. The red dashed line represents the number of genes identified by the Grace test with the KEGG network.} 
\label{fig:robust}
\centering
\includegraphics[width = 12cm]{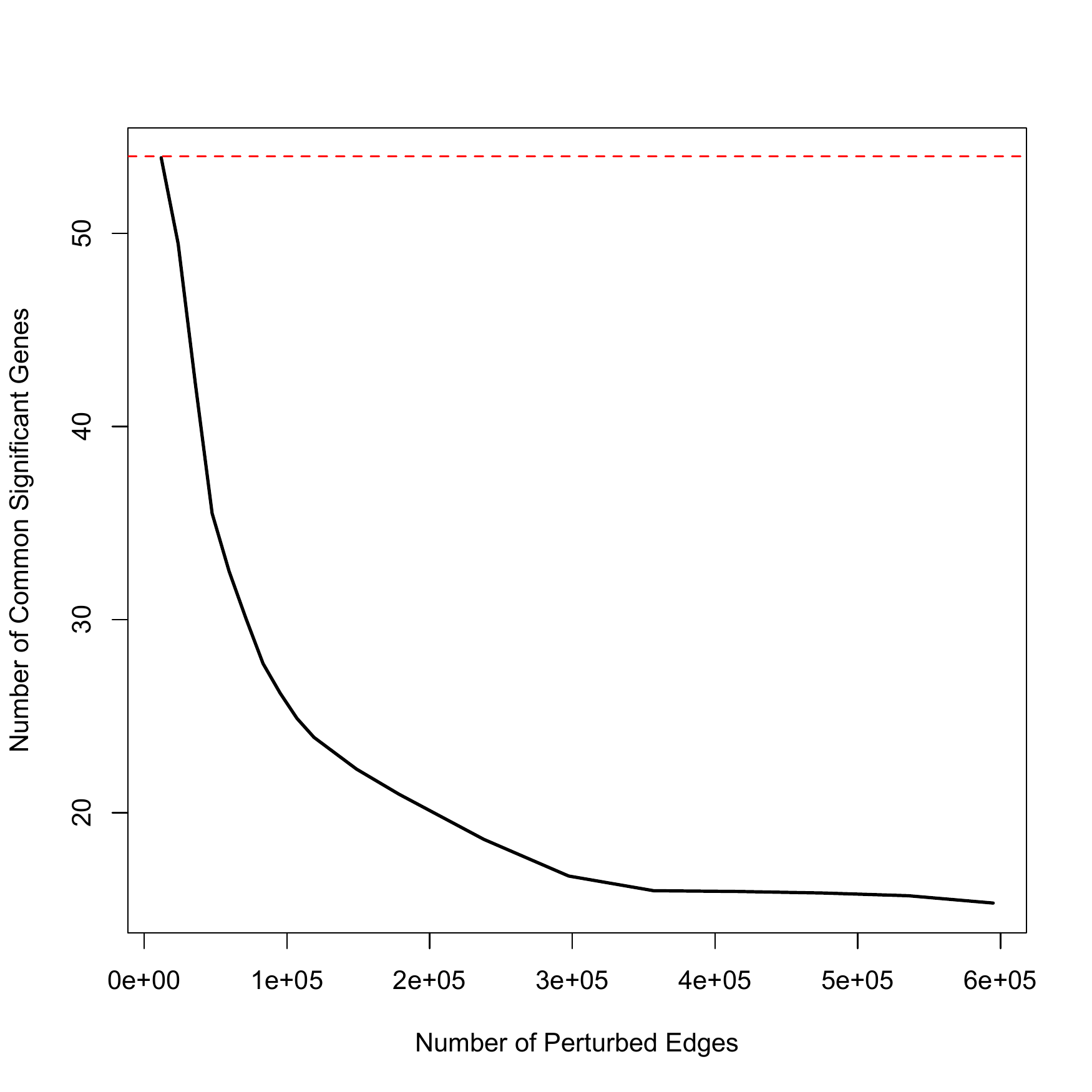} 
\end{figure}

We examine whether the result of the Grace test on the TCGA data is sensitive to the KEGG network structure. To this end, we randomly change the connectivity of $m$ node pairs in the KEGG network and form the new perturbed network $\tilde G$, $|E\Delta \tilde E|=m$, where $\Delta$ is the symmetric difference operator between two sets. In other words, for $m$ randomly selected node pairs $(a_i, b_i)$, $i=1,...,m$, if there is an edge $(a_i,b_i)$ in the KEGG network, we remove it in the perturbed network; otherwise, we add an edge in the perturbed network. In our examination, $m$ ranges from $10,000$ to $600,000$. Note that there are 38,541 edges in the original KEGG network. We counted the number of genes that are significant using both networks. The result shown in Figure~\ref{fig:robust} is an average of 50 independent replications.

\subsubsection{Prediction Performance}
We also compare the prediction performance by Grace, GraceR, GraceI and lasso with tuning parameters chosen by 10-fold CV, as well as ridge with $h_2=1$. The result is shown in Table~\ref{tab:RDpred}. GraceR produced the smallest CV prediction error, followed closely by GraceI and Grace. This result may indicate the KEGG network information is in fact informative in prediction. 
\begin{table}
\centering
\caption{Prediction performance of the Grace, GraceR, GraceI(ridge regression with tuning parameter chosen by CV), ridge ($h_2=1$) and lasso. The performance metric is the sum of 10-fold CV prediction error (CVER).} 
\label{tab:RDpred} 
\begin{tabular}{lccccc}
\hline \hline
	&Grace	&GraceR	&GraceI	&Ridge	&Lasso	\\	\hline
CVER	& 3473	& 3411	&3418	&3917	&3546\\
\hline
\end{tabular}
\end{table}
 
\subsection{Additional Simulation Studies with Extended NPE}

\begin{figure}
\caption{Comparison of power and type-I error rates of different testing methods with their 95\% confidence bands. Testing methods include LDPE, ridge, GraceI, Grace and GraceR. Filled circles ($\bullet$) show powers, whereas crosses ($\times$) are type-I error rates. Numbers on $x$-axis for Grace and GraceR tests refer to the number of perturbed edges (\textrm{NPE}).} 
\label{fig:power2}
\centering
\includegraphics[angle = 90, height = 20cm]{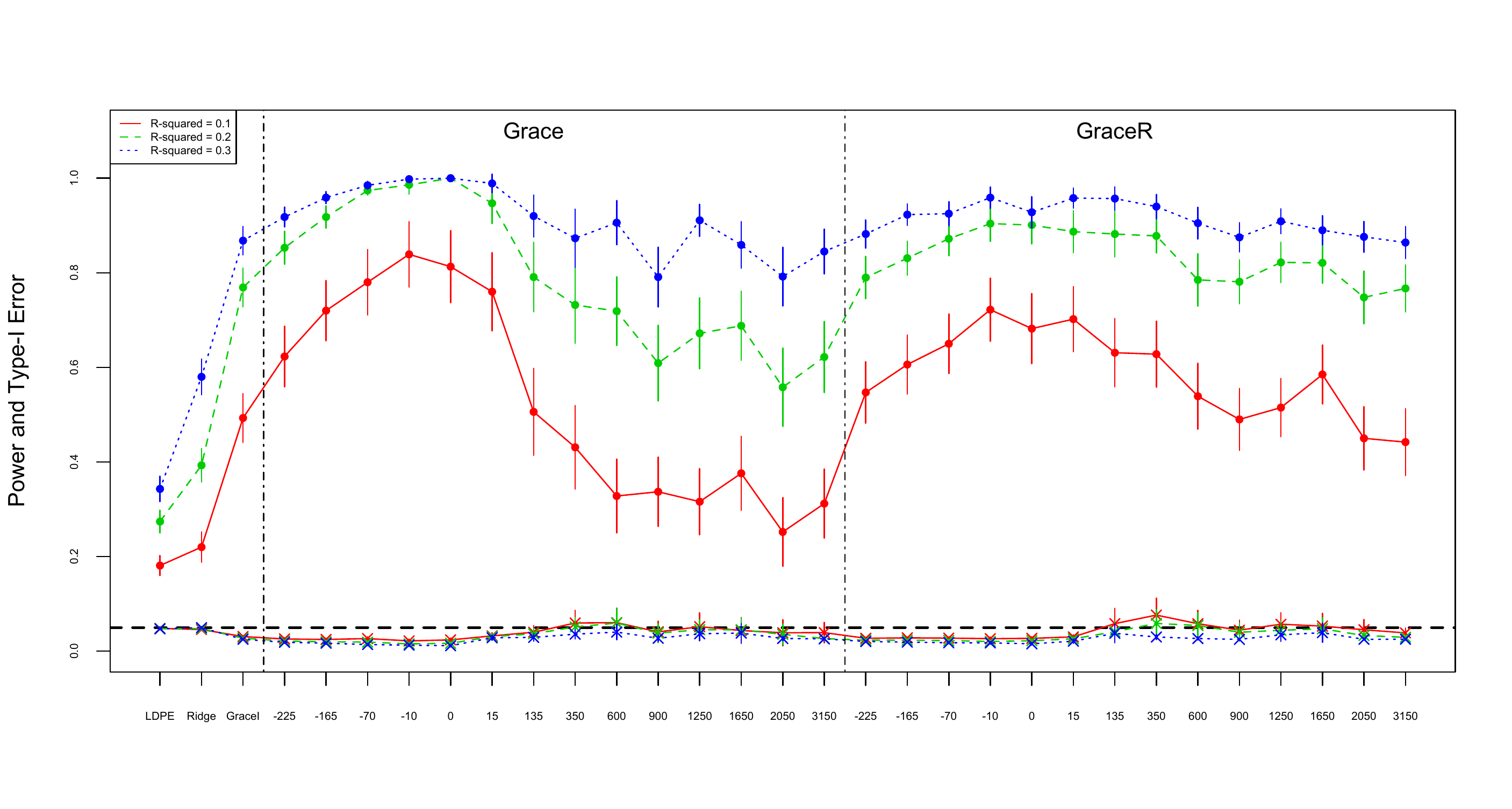} 
\end{figure}

We performed simulation studies with extended $\textrm{NPE}\in\{$\textsc{-225, -165, -70, -10, 0, 15, 135, 350, 600, 900, 1250, 1650, 2050, 3150}$\}$. These perturbations in the network correspond to the spectral norm of perturbations $\|\bL-\bL^*\|_2/\|\bL^*\|_2$ equal 0.85, 0.75, 0.50, 0.25, 0, 0.25, 0.50, 0.75, 1.00, 1.25, 1.50, 1.75, 2.00 and 2.65, respectively. The power and type-I error rates are summarized in Figure~\ref{fig:power2}, Table~\ref{tab:power} and Table~\ref{tab:level}. Our conclusions on the simulation study stated in the main paper do not change with this expanded version of simulation study.

\begin{table}
\centering
\caption{Mean power and the standard error for the LDPE test, ridge test, GraceI, Grace and GraceR tests with different $R^2$ values.} 
\label{tab:power} 
\begin{tabular}{lccc}
\hline \hline
             								& $R^2=0.1$ 		&$R^2=0.2$ 		&$R^2=0.3$   	\\	\hline
LDPE								& 0.181 (0.011)		& 0.274 (0.012)		& 0.343 (0.014) \\
Ridge        							& 0.220 (0.016)		& 0.393 (0.018) 	& 0.580 (0.019) \\
GraceI								& 0.493 (0.026)		& 0.769 (0.021) 	& 0.868 (0.015) \\
Grace $\mathrm{NPE} = \textsc{-225}$  		& 0.623 (0.033)		& 0.853 (0.018) 	& 0.918 (0.011) \\
Grace $\mathrm{NPE}  = \textsc{-165}$  		& 0.720 (0.032)		& 0.918 (0.012) 	& 0.959 (0.007) \\
Grace $\mathrm{NPE}  = \textsc{-70}$  		& 0.780 (0.035)		& 0.974 (0.005) 	& 0.985 (0.004) \\
Grace $\mathrm{NPE}  = \textsc{-10}$  		& 0.839 (0.035)		& 0.986 (0.010) 	& 0.998 (0.001) \\
Grace $\mathrm{NPE}  = \textsc{0}$  		& 0.813 (0.039)		& 1.000 (0.000) 	& 1.000 (0.000) \\
Grace $\mathrm{NPE}  = \textsc{15}$  		& 0.760 (0.042)		& 0.947 (0.022) 	& 0.989 (0.010) \\
Grace $\mathrm{NPE}  = \textsc{135}$  		& 0.506 (0.047)		& 0.791 (0.038) 	& 0.920 (0.023) \\
Grace $\mathrm{NPE}  =\textsc{350}$  		& 0.431 (0.045)		& 0.732 (0.041) 	& 0.873 (0.031) \\
Grace $\mathrm{NPE}  = \textsc{600}$  		& 0.328 (0.040)		& 0.719 (0.037) 	& 0.906 (0.024) \\
Grace $\mathrm{NPE}  = \textsc{900}$	  	& 0.337 (0.037)		& 0.609 (0.041) 	& 0.791 (0.032) \\
Grace $\mathrm{NPE}  = \textsc{1250}$  		& 0.316 (0.036)		& 0.672 (0.038) 	& 0.911 (0.017) \\
Grace $\mathrm{NPE}  = \textsc{1650}$  		& 0.376 (0.040)		& 0.688 (0.037) 	& 0.859 (0.025) \\
Grace $\mathrm{NPE}  = \textsc{2050}$  		& 0.252 (0.037)		& 0.558 (0.042) 	& 0.792 (0.032) \\
Grace $\mathrm{NPE}  = \textsc{3150}$  		& 0.312 (0.037)		& 0.622 (0.038) 	& 0.845 (0.024) \\
GraceR $\mathrm{NPE} = \textsc{-225}$  		& 0.547 (0.033)		& 0.790 (0.023) 	& 0.882 (0.015) \\
GraceR $\mathrm{NPE} = \textsc{-165}$  		& 0.606 (0.032)		& 0.831 (0.018) 	& 0.923 (0.012) \\
GraceR $\mathrm{NPE}  = \textsc{-70}$  		& 0.650 (0.032)		& 0.872 (0.018) 	& 0.925 (0.013) \\
GraceR $\mathrm{NPE}  = \textsc{-10}$  		& 0.722 (0.034)		& 0.904 (0.019) 	& 0.959 (0.011) \\
GraceR $\mathrm{NPE}  = \textsc{0}$  		& 0.682 (0.038)		& 0.901 (0.020) 	& 0.928 (0.017) \\
GraceR $\mathrm{NPE}  = \textsc{15}$  		& 0.702 (0.035)		& 0.887 (0.023) 	& 0.958 (0.011) \\
GraceR $\mathrm{NPE}  = \textsc{135}$  		& 0.631 (0.037)		& 0.882 (0.025) 	& 0.957 (0.013) \\
GraceR $\mathrm{NPE}  = \textsc{350}$  		& 0.628 (0.036)		& 0.878 (0.018) 	& 0.940 (0.013) \\
GraceR $\mathrm{NPE}  = \textsc{600}$  		& 0.539 (0.036)		& 0.785 (0.028) 	& 0.905 (0.017) \\
GraceR $\mathrm{NPE}  = \textsc{900}$		& 0.490 (0.033)		& 0.781 (0.024) 	& 0.875 (0.016) \\
GraceR $\mathrm{NPE}  = \textsc{1250}$  	& 0.515 (0.031)		& 0.822 (0.022) 	& 0.909 (0.013) \\
GraceR $\mathrm{NPE}  = \textsc{1650}$  	& 0.585 (0.032)		& 0.821 (0.022) 	& 0.890 (0.016) \\
GraceR $\mathrm{NPE}  = \textsc{2050}$  	& 0.450 (0.034)		& 0.748 (0.028) 	& 0.876 (0.017) \\
GraceR $\mathrm{NPE}  = \textsc{3150}$  	& 0.442 (0.036)		& 0.767 (0.025) 	& 0.864 (0.017) \\
\hline
\end{tabular}
\end{table}

\begin{table}
\centering
\caption{Mean type-I error rate and the standard error for the LDPE test, ridge test, GraceI, Grace and GraceR tests with different $R^2$ values.} 
\label{tab:level} 
\begin{tabular}{lccc}
\hline \hline
             								& $R^2=0.1$ 		&$R^2=0.2$ 		&$R^2=0.3$ 	\\	\hline
LDPE								& 0.048 (0.0010)	& 0.048 (0.0010)	& 0.047 (0.0010) \\
Ridge        							& 0.046 (0.0012)	& 0.048 (0.0013) 	& 0.050 (0.0012) \\
GraceI 								& 0.031 (0.0010)	& 0.027 (0.0009) 	& 0.025 (0.0008) \\
Grace $\mathrm{NPE} = \textsc{-225}$  		& 0.026 (0.0013)	& 0.021 (0.0012) 	& 0.019 (0.0010) \\
Grace $\mathrm{NPE} = \textsc{-165}$  		& 0.025 (0.0014)	& 0.020 (0.0013) 	& 0.017 (0.0012) \\
Grace $\mathrm{NPE} = \textsc{-70}$  		& 0.027 (0.0021)	& 0.019 (0.0017) 	& 0.014 (0.0013) \\
Grace $\mathrm{NPE} = \textsc{-10}$  		& 0.022 (0.0021)	& 0.015 (0.0017) 	& 0.013 (0.0015) \\
Grace $\mathrm{NPE} = \textsc{0}$  			& 0.024 (0.0021)	& 0.017 (0.0017) 	& 0.011 (0.0013) \\
Grace $\mathrm{NPE} = \textsc{15}$  		& 0.032 (0.0034)	& 0.031 (0.0031) 	& 0.028 (0.0028) \\
Grace $\mathrm{NPE} = \textsc{135}$  		& 0.040 (0.0073)	& 0.037 (0.0059) 	& 0.029 (0.0042) \\
Grace $\mathrm{NPE} = \textsc{350}$  		& 0.059 (0.0137)	& 0.051 (0.0102) 	& 0.036 (0.0052) \\
Grace $\mathrm{NPE} = \textsc{600}$  		& 0.060 (0.0156)	& 0.059 (0.0155) 	& 0.040 (0.0083) \\
Grace $\mathrm{NPE} = \textsc{900}$	  	& 0.041 (0.0115)	& 0.038 (0.0101) 	& 0.027 (0.0033) \\
Grace $\mathrm{NPE} = \textsc{1250}$  		& 0.052 (0.0151)	& 0.045 (0.0111) 	& 0.037 (0.0075) \\
Grace $\mathrm{NPE} = \textsc{1650}$  		& 0.044 (0.0141)	& 0.045 (0.0125) 	& 0.038 (0.0104) \\
Grace $\mathrm{NPE} = \textsc{2050}$  		& 0.039 (0.0141)	& 0.035 (0.0112) 	& 0.027 (0.0023) \\
Grace $\mathrm{NPE} = \textsc{3150}$  		& 0.039 (0.0110)	& 0.027 (0.0024) 	& 0.026 (0.0015) \\
GraceR $\mathrm{NPE} = \textsc{-225}$  		& 0.027 (0.0012)	& 0.023 (0.0011) 	& 0.020 (0.0009) \\
GraceR $\mathrm{NPE} = \textsc{-165}$  		& 0.028 (0.0013)	& 0.023 (0.0011) 	& 0.019 (0.0010) \\
GraceR $\mathrm{NPE} = \textsc{-70}$  		& 0.028 (0.0014)	& 0.022 (0.0014) 	& 0.018 (0.0012) \\
GraceR $\mathrm{NPE} = \textsc{-10}$  		& 0.026 (0.0018)	& 0.020 (0.0015) 	& 0.017 (0.0014) \\
GraceR $\mathrm{NPE} = \textsc{0}$  		& 0.027 (0.0018)	& 0.022 (0.0016) 	& 0.015 (0.0013) \\
GraceR $\mathrm{NPE} = \textsc{15}$  		& 0.030 (0.0025)	& 0.026 (0.0025) 	& 0.021 (0.0025) \\
GraceR $\mathrm{NPE} = \textsc{135}$  		& 0.058 (0.0165)	& 0.041 (0.0112) 	& 0.038 (0.0103) \\
GraceR $\mathrm{NPE} = \textsc{350}$  		& 0.076 (0.0182)	& 0.059 (0.0152) 	& 0.030 (0.0027) \\
GraceR $\mathrm{NPE} = \textsc{600}$  		& 0.058 (0.0145)	& 0.054 (0.0139) 	& 0.027 (0.0016) \\
GraceR $\mathrm{NPE} = \textsc{900}$		& 0.044 (0.0109)	& 0.040 (0.0099) 	& 0.025 (0.0010) \\
GraceR $\mathrm{NPE} = \textsc{1250}$  	& 0.057 (0.0125)	& 0.044 (0.0100) 	& 0.034 (0.0071) \\
GraceR $\mathrm{NPE} = \textsc{1650}$  	& 0.053 (0.0138)	& 0.047 (0.0122) 	& 0.039 (0.0104) \\
GraceR $\mathrm{NPE} = \textsc{2050}$  	& 0.045 (0.0111)	& 0.033 (0.0038) 	& 0.025 (0.0009) \\
GraceR $\mathrm{NPE} = \textsc{3150}$  	& 0.039 (0.0053)	& 0.029 (0.0017) 	& 0.025 (0.0012) \\
\hline
\end{tabular}
\end{table}

\label{lastpage}
\bibliographystyle{apa}
\bibliography{NCPRpaper}
\end{document}